\documentclass[12pt]{article}
\usepackage[english,russian]{babel}
\usepackage[cp1251]{inputenc}%
\usepackage{graphicx}
\usepackage{epsfig}
\usepackage{ulem}
\usepackage[dvipsnames]{color}
 \title{Traveling surface wave propagation on shallow water with variable
 bathymetry and current}
 \author{Semyon Churilov (Семён Чурилов),\\
 {\footnotesize\it Institute of Solar-Terrestrial Physics of the Siberian
 Branch of Russian Academy of Sciences,}\\
 {\footnotesize\it 126a Lermontov St., Irkutsk, 664033, Russia.}}%
 \date{}
\begin{document}
\newcommand{\ba}{\begin{array}}
\newcommand{\ea}{\end{array}}
\newcommand{\be}{\begin{equation}}
\newcommand{\ee}{\end{equation}}
\newcommand{\bea}{\begin{eqnarray}}
\newcommand{\eea}{\end{eqnarray}}
\newcommand{\bfig}{\begin{figure}}
\newcommand{\efig}{\end{figure}}
\newcommand{\Bl}{\Bigl}
\newcommand{\Br}{\Bigr}
\newcommand{\RE}{{\rm Re}\,}
\newcommand{\IM}{{\rm Im}\,}
\newcommand{\re}{{\rm e}}
\newcommand{\bA}{\bar{A}}
\newcommand{\bB}{\bar{B}}
\newcommand{\bvp}{\bar{\varphi}}
\newcommand{\bP}{\bar{\Phi}}
\newcommand{\cD}{{\cal D}}
\newcommand{\cF}{{\cal F}}
\newcommand{\cH}{{\cal H}}
\newcommand{\hB}{\hat{B}}
\newcommand{\hC}{\hat{C}}
\newcommand{\hx}{\hat{x}}
\newcommand{\tu}{\tilde{u}}
\newcommand{\tG}{\tilde{G}}
\newcommand{\tI}{\tilde{I}}
\newcommand{\ptl}{\partial}
\newcommand{\dd}{{\rm d}}
\newcommand{\al}{\alpha}
\newcommand{\Gm}{\Gamma}
\newcommand{\Lb}{\Lambda}
\newcommand{\lb}{\lambda}
\newcommand{\vp}{\varphi}
\newcommand{\vep}{\varepsilon}
\newcommand{\om}{\omega}
\newcommand{\Om}{\Omega}
\newcommand{\sg}{\sigma}
\newcommand{\tB}{\tilde{B}}
\newcommand{\tV}{\tilde{V}}
\newcommand{\tW}{\tilde{W}}
\newcommand{\tZ}{\tilde{Z}}
\newcommand{\din}{\displaystyle\int\limits}
\newcommand{\IN}{\displaystyle\int\limits_{0}^}
\newcommand{\II}{\displaystyle\int\limits^{\infty}_}
\newcommand{\IE}{\displaystyle\int\limits^{1}_}
\newcommand{\dfrac}{\displaystyle\frac}
\renewcommand{\abstractname}{}
\renewcommand{\thesection}{\arabic{section}}
\renewcommand{\thesubsection}{\arabic{section}.\arabic{subsection}}
\renewcommand{\thesubsubsection}{\arabic{section}.\arabic{subsection}.%
\arabic{subsubsection}}
 \maketitle
 \vspace{-10mm}
 {\selectlanguage{english}
 \begin{abstract}\noindent
 Energy transmission over long distances by waves is a key mechanism for
 many natural processes. This possibility arises when an inhomogeneous medium
 is arranged in such a manner that it enables a certain type of wave to
 propagate with virtually no reflection or scattering. By application of the
 Laplace cascade method for integrating second-order hyperbolic equations, a
 general algorithm for finding the parameters of inhomogeneous non-reflecting
 flows is proposed. The algorithm is applied to the problem of long linear
 surface waves propagation in a channel with variable cross-section. The
 general analysis of the problem is illustrated by a few representative
 solutions and compared with the results of previous studies. The results
 obtained may be of interest to mitigate the possible impact of waves on
 ships, marine engineering constructions, and human coastal activities.
 \end{abstract}
 \section{Introduction}
 \label{sec1}
 \hspace\parindent
 Wave propagation in an inhomogeneous medium is typically accompanied
 by reflection and scattering which significantly attenuate the directed transfer
 of energy. It is well known, however, that in an isothermal atmosphere,where
 the speed of sound is constant and the density and pressure decrease with
 height $z$ as $\exp(-z/H_0)$, acoustic-gravity waves propagate without
 reflection \cite{Gossard}, that is, they are {\it traveling waves}. This
 property also holds for some other dependences of the speed of sound on $z$
 \cite{Petr11}. Another example. A theoretical analysis of surface wave
 propagation in the coastal zone revealed that there is no reflection if the
 depth $H$ varies with distance $x$ from the shore line as $x^{4/3}$.
 And this is precisely the relationship obtained by depth measurements in
 Pirita Bay (Estonia) \cite{Soom1,Soom2}.

 Today, a vast literature is devoted to the search for configurations of
 various inhomogeneous media that allow the propagation of traveling waves.
 In the course of these investigations, the very concept of what a traveling
 wave should look like has significantly changed. Initially it was based
 on the classical equation for waves propagating with the velocity $c$,
 \be
 \dfrac{\ptl^2 f}{\ptl t^2} - c^2\dfrac{\ptl^2 f}{\ptl x^2} = 0,
 \label{WE}
 \ee
 whose general solution in the case $c=\mbox{const}$,
 \be
 f(x,t) = F_1\left(t-\dfrac{x}{c}\right) + F_2\left(t+\dfrac{x}{c}\right),
 \label{TW0}
 \ee
 consists of two waves of arbitrary shape propagating in opposite directions,
 as well as on its solution in the Wentzel--Kramers--Brillouin approximation
 \cite{Heading,Froman} for a monochromatic perturbation in a medium with
 $c=c(x)$,
 \[
 f(x,\,t) = \sqrt{c(x)}\Bl\{C_1\exp\Bl[-i\om(t - \hx)\Br] +
 C_2\exp\Bl[-i\om(t + \hx)\Br]\Br\}, \qquad
 \hx = \!\int\dfrac{\dd x}{c(x)}\,,
 \]
 where $\om$ is the frequency, and $C_{1,2}=\mbox{const}$ are the relative
 amplitudes of waves.

 For a long time, the attention of researchers was focused on waves
 in media at rest, and the main method was in using a substitution
 $f(x,\,t) = a(x)f_1(\hx,\,t)$ to reduce the wave equation with $x$-dependent
 coefficients to some reference equation with constant coefficients for the function
 $f_1(\hx,\,t)$, for example, to Eq.~(\ref{WE}). This yields a solution of
 the form
 \be
 f(x,t) = A(x)\Bl[F_1(t-\hx) + F_2(t+\hx)\Br]
 \label{TW1}
 \ee
 differing from Eq.~(\ref{TW0}) only in the form-factor $A(x)$
 which is common to both waves.

 Recently, the set of reference equations used has been significantly extended
 \cite{Kap22,Kap23,Petr23} by including the Euler--Darboux--Poisson equation
 with integer $m$,
 \be
 \dfrac{\ptl^2 f}{\ptl t^2} - \dfrac{\ptl^2 f}{\ptl\hx^2} - \dfrac{2m}{\hx}\,
 \dfrac{\ptl f}{\ptl\hx} = 0
 \label{EDP}
 \ee
 and related equations with variable coefficients. For $m=0$, Eq.~(\ref{EDP})
 is equivalent to Eq.~(\ref{WE}), and for $m=1$ it describes
 spherically symmetric traveling waves in a homogeneous three-dimensional
 medium. For other integer $m$, its solution also describes two independent
 traveling waves, but of a more complex structure, each as a
 finite sum (see, e.g., \cite{Mises}, \S 12.4),
 \be
 f(\hx,t) = \sum_{k=0}^{|m|}C_{k;m}\hx^{k+m_0}\dfrac{\ptl^k F}{\ptl\hx^k},
 \qquad F(\hx,t) = F_1(t-\hx) + F_2(t+\hx), \quad C_{k;m}=\mbox{const},
 \label{TW}
 \ee
 where $m_0=1-2m$ for $m>0$, $m_0=0$ for $m<0$, $C_{0;m}=1$, and the
 remaining constants are easily found by substituting into Eq.~(\ref{EDP}).
 Here, as in solution (\ref{TW1}), each wave has its own
 arbitrary {\it phase function}, but the waveform depends not only on it,
 but also on its derivatives. In other words, $F_1$ serves as
 the generating function for all functions of $(t-\hx)$ that describe the
 structure of the first wave, and $F_2$ serves as the generating function
 for all functions of $(t+\hx)$ that describe the second wave.

 As a result, a more general concept of the traveling wave structure was
 formed as compared with Eq.~(\ref{TW1}). It became clear that it can be
 described by a finite sum of the form
 \be
 f(x,t) = \sum_{k=0}^{n}A_k(x)\Psi_k(T_c), \quad T_c = t-\hx,
 \label{TW2}
 \ee
 where $n\ge 0$, the form-factors $A_k(x)$ are determined by properties of
 the medium, and the phase functions $\Psi_k(T_c)$ have a common
 generating function $\Psi(T_c)$. In an isotropic medium, a traveling wave
 propagating in the opposite direction is described by a similar sum with
 the same form-factors but its own generating function $\tilde\Psi(t+\hx)$.
 In an anisotropic medium, waves traveling in opposite
 directions typically have different form-factors.

 If an inhomogeneous medium moves with a time-independent velocity $U(x)$,
 it remains stationary but the waves are carried away by the current, and
 their velocities $w_{1,2}(x)=U(x)\pm c(x)$ are {\it different} functions
 of $x$. Taking this into account, another approach was proposed to searching
 for non-reflecting flows based on factorization of the wave equation and
 finding its solutions of the form (\ref{TW1}). It was formulated
 in papers \cite{ChSt22,ChSt22a} (hereinafter referred to as Paper I and
 Paper II), devoted to the problem of surface waves propagation in a channel
 of variable cross-section, and then was adapted to internal waves in flows
 of a shallow two-layer medium \cite{Chur23,Chur23a} and Alfv\'{e}n waves
 in plasma flows along a magnetic field \cite{Chur24a}. Comprehending
 the results obtained has shown that the factorization idea is
 fruitful, but searching for a solution in the form (\ref{TW1}) allows us to
 identify only a part of the non-reflecting flows. For this reason, we
 return to the problem of surface wave propagation in channels to provide
 a more general algorithm for finding non-reflecting flows, based on the
 representation (\ref{TW2}) for a traveling wave.

 In the theory of integration of second-order linear hyperbolic equations,
 the Laplace cascade method improved by Legendre and Imshenetsky (see
 \cite{Imsh,Goursat,Darboux} and, in a more modern version, \cite{Meleshko}),
 has been known for over two centuries. It is based on transforming the
 dependent variable according to a specific algorithm. The key result for us
 is that a solution of the form (\ref{TW2}) exists if and only if the wave
 equation factorizes after a series (cascade) of transformations. It turns
 out that $\Psi_k$ and $\Psi$ in such a solution are related in the same
 manner as in Eq.~(\ref{TW}),
 \be
 f(x,t) = A_0(x)\Psi(T_c) + \sum\limits_{k=1}^{n}A_k(x)
 \dfrac{\dd^k\Psi}{\dd T_c^k}\,.
 \label{TWF}
 \ee

 For the initial wave equation to be factorized at some ($n$-th) step of the
 transformation cascade, the equation coefficients expressed in terms of the
 flow parameters must satisfy certain conditions (specific for each $n$)
 which distinguish non-reflecting flows from the set of possible flows. In
 this paper, we construct an algorithm for finding non-reflection conditions
 and form-factors $A_k$ for any $n$, based on the representation (\ref{TWF})
 for a traveling wave. Within the approach adopted in Papers I and II, this
 was done only for $n=0$ (flows of class $A$) and, in part, for $n=1$ (flows
 of classes $B$ and $C$). This paper is a further development of Papers I
 and II, and we will refer to the figures and equations given there prefixing
 their numbers with I or II (e.g., Figure I.3 or Eq.~(II.2.22)).

 The text is organized as follows. In Section~\ref{sec2}, the basic equations
 are derived, and the cases $n=0$ and $n=1$ are considered in Section~\ref{sec3}.
 The non-reflectivity conditions for $n=2$ and the flows satisfying them are
 studied in Section~\ref{sec4}. The results obtained are discussed from the
 physical point of view in Section~\ref{sec5}, and Section~\ref{sec6}
 contains conclusions. Appendix~\ref{AppA} provides a brief description
 of the Laplace cascade method as applied to the problem under consideration,
 and Appendix~\ref{AppB} is devoted to solving Eq.~(\ref{Eq-y}).
 \section{Basic equations}
 \label{sec2}
 \hspace*{\parindent}
 Consider a steady flow of shallow water in a channel with a
 width $D(x)$, a bottom profile $z_B = B(x)$, and a flow velocity $U(x)$
 varying along the flow  direction (see Figure~\ref{fg1}).
%
 \begin{figure}[!h]
 \epsfxsize=100mm
 \centerline{\epsfbox{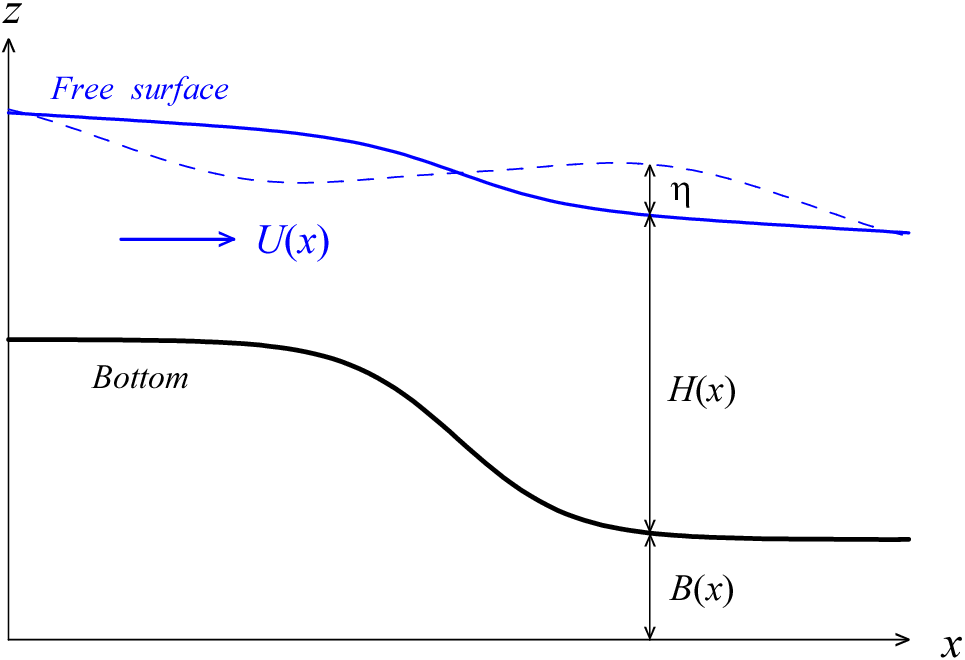}}
 \vspace{.2cm}
 \caption{Sketch of the flow configuration in the vertical plane. The
 perturbed surface is shown by dashes.}
 \label{fg1}
 \end{figure}
 The mass conservation,
 \be
 U(x)H(x)D(x) = {\rm const},
 \label{Flux}
 \ee
 and the Bernoulli equation,
 \be
 \frac{1}{2}\,U^2(x) + \hat{g}\Bl(H(x) + B(x)\Br) = {\rm const},
 \label{Bern}
 \ee
 provide independent variation along the longitudinal channel $x$-axis of
 the flow, $U(x)$, and wave, $c(x)=\sqrt{\hat{g}H(x)}$, velocities which are
 assumed to be positive everywhere. Here  $H(x)$ is the flow depth, and
 $\hat{g}$ is the acceleration due to gravity.

 In the shallow water approximation, linear waves are described by the
 Euler equations,
 \be
 \dfrac{\ptl u}{\ptl t}+\dfrac{\ptl(Uu)}{\ptl x} =
 -\dfrac{\ptl\zeta}{\ptl x}, \quad \zeta = \hat{g}\eta,
 \label{Euler}
 \ee
 and mass balance equation,
 \[
 \dfrac{\ptl\eta}{\ptl t} + \dfrac{1}{D}\,\dfrac{\ptl}{\ptl x}
 \Bl[D(U\,\eta + H u)\Br] = 0,
 \]
 which, in view of Eq.~(\ref{Flux}), can be conveniently written as
 \be
 \dfrac{\ptl\zeta}{\ptl t} + c^2U\dfrac{\ptl}{\ptl x}\left(\dfrac{\zeta}{c^2}
 + \dfrac{u}{U}\right) = 0.
 \label{eta}
 \ee
 Here $u(x,t)$ and $\eta(x,t)$ are the wave-induced velocity disturbance
 and the deviation of the free surface from the equilibrium. Equations
 (\ref{Euler}) and (\ref{eta}) form a linear hyperbolic system with
 characteristics
 \be
 \dfrac{\dd x}{\dd t} = U(x) \pm c(x),
 \label{har}
 \ee
 corresponding to the propagation of accelerated ($+$) and retarded ($-$)
 waves.

 Let us assume that the flow parameters (i.e., the velocities $U$ and $c$)
 are such that the waves propagate without reflection. Following
 Eq.~(\ref{TWF}), we seek solutions to Eqs.~(\ref{Euler}) and (\ref{eta})
 for the accelerated wave in the form
 \be
 \ba{l}
 u_+(x,t) = V_0(x)F_1(T_+) + \sum\limits_{m=1}^{N} V_m(x)F_1^{(m)}(T_+),
   \\ \\
 \zeta_+(x,t) = Z_0(x)F_1(T_+) + \sum\limits_{m=1}^{N} Z_m(x)F_1^{(m)}(T_+),
 \ea
 \label{+}
 \ee
 and for the retarded wave -- in the form
 \be
 \ba{l}
 u_-(x,t) = \tV_0(x)F_2(T_-) + \sum\limits_{m=1}^{n} \tV_m(x)F_2^{(m)}(T_-),
   \\ \\
 \zeta_-(x,t) = \tZ_0(x)F_2(T_-) + \sum\limits_{m=1}^{n}\tZ_m(x)F_2^{(m)}(T_-),
 \ea
 \label{-}
 \ee
 where $F_{1,\,2}(X)$ are arbitrary functions,
 \[
 F_{1,\,2}^{(m)}(X) = \dfrac{\dd^m F_{1,\,2}}{\dd X^m}, \qquad
 T_\pm = t-\int\dfrac{\dd x}{U(x) \pm c(x)}.
 \]
 Due to the linearity of the problem, the functions $V_i(x)$ and $Z_i(x)$
 are defined up to a common numerical factor which will be chosen later. The
 same is also true for $\tV_i(x)$ and $\tZ_i(x)$.

 Substituting the expansions (\ref{+}) into Eqs.~(\ref{Euler}) and
 (\ref{eta}) yields (hereinafter the prime denotes the derivative with
 respect to the function argument)
 \[
 \ba{l}
 (UV_0+Z_0)'F_1 + \dfrac{cV_0-Z_0}{U+c}\,F'_1 + \sum\limits_{m=1}^{N}\left[
 (UV_m+Z_m)'F_1^{(m)} + \dfrac{cV_m-Z_m}{U+c}\,F_1^{(m+1)}\right] = 0,
   \\ \\
 c^2U\left(\dfrac{Z_0}{c^2} + \dfrac{V_0}{U}\right)'F_1 -
 \dfrac{cV_0-Z_0}{U+c}\,cF'_1 +
 \sum\limits_{m=1}^{N}\left[c^2U\left(\dfrac{Z_m}{c^2} +
 \dfrac{V_m}{U}\right)'F_1^{(m)}-\dfrac{cV_m-Z_m}{U+c}\,
 cF_1^{(m+1)}\right] = 0.
 \ea
 \]
 Since $F_1(T_+)$ is arbitrary, the coefficients at it and its derivatives
 must vanish. Therefore
 \be
 \ba{l}
 \dfrac{\dd}{\dd x}\left(\dfrac{V_0}{U} + \dfrac{Z_0}{c^2}\right) =
 \dfrac{\dd}{\dd x}\Bl(UV_0+Z_0\Br) = 0\ \ \Longrightarrow\ \
 V_0 = \dfrac{B_1c^2-B_2}{c^2-U^2}\,U, \quad
 Z_0 = \dfrac{B_2-B_1U^2}{c^2-U^2}\,c^2,
   \\ \\
 c\,U\dfrac{\dd}{\dd x}\left(\dfrac{V_{m+1}}{U} + \dfrac{Z_{m+1}}{c^2}\right)
 = -\dfrac{\dd}{\dd x}\Bl(UV_{m+1}+Z_{m+1}\Br) = \dfrac{cV_m-Z_m}{U+c}\,,
 \ \ 0\le m\le N-1,
    \\ \\
 Z_N = cV_N\ \ \stackrel{m=N-1}{\Longrightarrow}\ \ \dfrac{\dd}{\dd x}
 \dfrac{(U+c)^2V_N^2}{c\,U} = 0\ \ \Longrightarrow\ \
 V_N(x) = \dfrac{a(x)}{U(x)+c(x)}\,, \quad a=(c\,U)^{1/2},
 \ea
 \label{Eq+}
 \ee
 where $B_1$ and $B_2$ are arbitrary constants, and the choice of unity as
 the constant of the last integration provides the normalization of
 $V_i(x)$ and $Z_i(x)$. The equations for the retarded wave are derived in
 the same way,
 \be
 \ba{l}
 \dfrac{\dd}{\dd x}\left(\dfrac{\tV_0}{U} + \dfrac{\tZ_0}{c^2}\right) =
 \dfrac{\dd}{\dd x}\Bl(U\tV_0+\tZ_0\Br) = 0\ \ \Longrightarrow\ \
 \tV_0 = \dfrac{\tB_1c^2-\tB_2}{c^2-U^2}\,U, \quad
 \tZ_0 = \dfrac{\tB_2-\tB_1U^2}{c^2-U^2}\,c^2,
   \\ \\
 c\,U\dfrac{\dd}{\dd x}\left(\dfrac{\tV_{m+1}}{U}+\dfrac{\tZ_{m+1}}{c^2}\right)
 = \dfrac{\dd}{\dd x}\Bl(U\tV_{m+1}+\tZ_{m+1}\Br) = \dfrac{c\tV_m+\tZ_m}{U-c},
 \ \ 0\le m\le n-1,
   \\ \\
 \tZ_n + c\tV_n = 0\ \ \stackrel{m=n-1}{\Longrightarrow}\ \
 \dfrac{\dd}{\dd x}\dfrac{(U-c)^2\tV_n^2}{c\,U} = 0\ \ \Longrightarrow\ \
 \tV_n(x) = \dfrac{a(x)}{U(x)-c(x)}\,.
 \ea
 \label{Eq-}
 \ee
 Sometimes, instead of $Z_i(x)$ and $\tZ_i(x)$, it is convenient to use the
 functions $W_i(x)$ and $\tW_i(x)$,
 \[
 Z_i(x) = c(x)\Bl[V_i(x) - W_i(x)\Br], \quad
 \tZ_i(x) = -c(x)\Bl[\tV_i(x) - \tW_i(x)\Br],
 \]
 satisfying the equations ($0\le m < N$)
 \be
 \ba{l}
 a^2\dfrac{\dd}{\dd x}\,\dfrac{(U+c)V_{m+1}-UW_{m+1}}{a^2} =
 -\dfrac{\dd}{\dd x}\Bl[(U+c)V_{m+1}-cW_{m+1}\Br] = \dfrac{c\,W_m}{U+c}\,,
   \\ \\
 W_0 = \dfrac{B_1a^2-B_2}{c-U}\,, \quad W_N = 0,
 \ea
 \label{Eq+1}
 \ee
 and ($0\le m < n$)
 \be
 \ba{l}
 a^2\dfrac{\dd}{\dd x}\,\dfrac{(U-c)\tV_{m+1}-U\tW_{m+1}}{a^2} =
 -\dfrac{\dd}{\dd x}\Bl[(U-c)\tV_{m+1}+c\,\tW_{m+1}\Br] =
 -\dfrac{c\,\tW_m}{U-c}\,,
   \\ \\
 \tW_0 = \dfrac{\tB_1a^2+\tB_2}{U+c}\,, \quad \tW_n = 0.
 \ea
 \label{Eq-1}
 \ee
 Equations (\ref{Eq-}) and (\ref{Eq-1}) for retarded waves differ from
 Eqs.~(\ref{Eq+}) and (\ref{Eq+1}) by tildes and a formal replacement of
 $N$ with $n$, $c$ with $-c$, and $a^2=c\,U$ with $-a^2$.

 We are seeking solutions of a prescribed form, so the
 systems (\ref{Eq+}) and (\ref{Eq-}) are overdetermined: in each, the
 number of equations is one greater than the number of unknown functions.
 For them to be solvable, the velocities $U(x)$ and $c(x)$ must satisfy a
 certain condition, specific for each $N\ (n)$, which defines {\it the
 set of non-reflecting flows of rank} $N\ (n)$. From physical considerations,
 it is clear that for $n=N$ these sets are the same, and we shall see below
 that this is true.

 For every $N$ there is its own solvability condition, and therein lies the
 meaning of the concept of rank. It should be kept in mind, however, that
 flows of a given rank are a subset of flows of any higher rank. Indeed, the
 rank of a flow is determined by the number of terms in the expansion
 (\ref{+}) or (\ref{-}). But it can easily be increased, for example, by
 substituting
 \[
 F_{1,2}(X) = G_{1,2}(X) + b_{1,2}^{(1)}\,\dfrac{\dd G_{1,2}}{\dd X} +
 b_{1,2}^{(2)}\,\dfrac{\dd^2 G_{1,2}}{\dd X^2} + \dots,
 \qquad b_{1,2}^{(i)}=\mbox{const}.
 \]
 \section{Currents of ranks 0 and 1}
 \label{sec3}
 \subsection{Currents of rank 0}
 \label{sec3-1}
 \hspace\parindent
 When $N=n=0$, the right-hand sides of Eqs.~(\ref{+}) and (\ref{-}) each
 contain only one term, and Eqs.~(\ref{Eq+1}) and (\ref{Eq-1})
 yield
 \[
 W_0(x) = 0\ \ \Longrightarrow\ \ a^2(x) = B_2/B_1, \quad
 \tW_0(x) = 0\ \ \Longrightarrow\ \ a^2(x) = \tB_2/\tB_1.
 \]
 Due to the arbitrariness of the constants $B_{1,2}$ and $\tB_{1,2}$,
 these conditions are equivalent to the common equation
 \be
 a(x) \equiv c(x)U(x) = a_* = \mbox{const},
 \label{R0}
 \ee
 and corresponding form-factors are
 \be
 \ba{l}
 V_0(x) = \dfrac{a_*}{U(x)+c(x)}\,, \quad Z_0(x) = c(x)V_0(x)\ \ \ \mbox{and}
   \\ \\
 \tV_0(x) = \dfrac{a_*}{U(x)-c(x)}\,, \quad \tZ_0(x) = c(x)\tV_0(x).
 \ea
 \label{form0}
 \ee
 Equation (\ref{R0}) describes a broad class of flows, referred to in Papers
 I and II as class $A$. One can specify the profile of one of the velocities,
 $c(x)$ or $U(x)$, along the entire $x$ axis as an arbitrary continuous
 positive function and, using Eq.~(\ref{R0}), obtain a family of corresponding
 profiles for the other velocity, ``numbered'' by the parameter $a_*$.
 Figure~\ref{fg2} shows some examples of rank 0 flows with $a_* = 1$.
%
 \begin{figure}[!h]
 \epsfxsize=100mm
 \centerline{\epsfbox{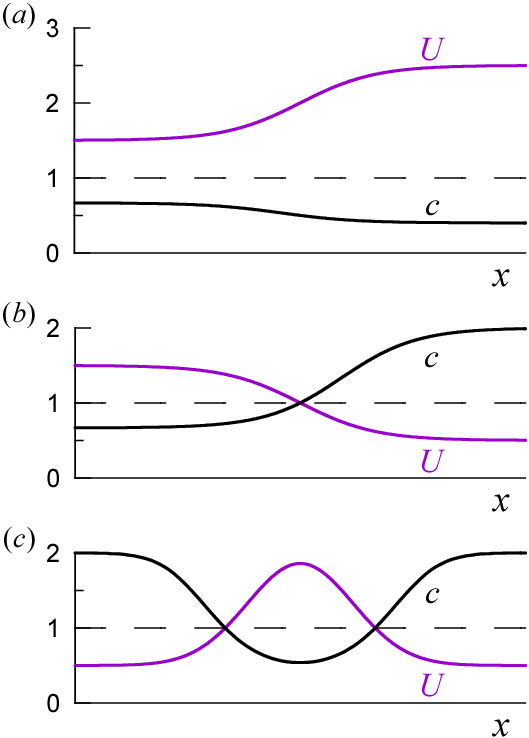}}
 \vspace{.2cm}
 \caption{Rank 0 flows,\ \,$c(x)\,U(x) = 1$.}
 \label{fg2}
 \end{figure}
 The $c(x)$ and $U(x)$ graphs can be interchanged. Then a subcritical flow
 ($U < c$) transforms into a supercritical one ($U > c$), and vice versa.
 It should be emphasized that these flows are regular in the critical
 point(s) where the flow velocity is equal to the wave velocity, $U(x)=c(x)$.
 In addition, it follows from Eqs.~(\ref{Flux}) and (\ref{R0}) that
 \be
 U(x)H^{1/2}(x) = \mbox{const}, \quad D(x)H^{1/2}(x) = \mbox{const}\ \
 \Longrightarrow\ \ U(x)/D(x) = \mbox{const},
 \label{UHD}
 \ee
 that is, the wider the channel, the higher the flow velocity.
 \subsection{Currents of rank 1}
 \label{sec3-2}
 \hspace\parindent
 Let us start with accelerated waves. From Eqs.~(\ref{Eq+}) and
 (\ref{Eq+1}), we find
 \[
 V_1 = \dfrac{a}{U+c}, \quad W_1 = 0, \quad
 \dfrac{\dd}{\dd x}\Bl[(U+c)V_1\Br] = -\dfrac{cW_0}{U+c}\,,
 \]
 and, after substituting $V_1$ and $W_0$ into the last equation, we obtain
 the solvability condition
 \be
 \dfrac{\dd a}{\dd x} = \dfrac{B_1a^2-B_2}{U^2-c^2}\,c.
 \label{Eq-a1}
 \ee
 Similarly, Eqs.~(\ref{Eq-}) and (\ref{Eq-1}) for retarded waves give
 \[
 \tV_1 = \dfrac{a}{U-c}, \quad \tW_1 = 0, \quad
 \dfrac{\dd}{\dd x}\Bl[(U-c)\tV_1\Br] = \dfrac{c\tW_0}{U-c}\,,
 \]
 or
 \be
 \dfrac{\dd a}{\dd x} = \dfrac{\tB_1a^2-\tB_2}{U^2-c^2}\,c,
 \label{Eq-a1-}
 \ee
 Since $B_{1,2}$ and $\tB_{1,2}$ are arbitrary constants, both
 solvability conditions define a common set of rank 1 flows for both wave
 types. We shall further describe it by Eq.~(\ref{Eq-a1}), the right-hand
 side of which is generally singular at the critical point(s) $U(x)=c(x)$.
 Its general solution depends on one arbitrary function (e.g., $c(x)$) and
 three arbitrary constants, including $B_1$ and $B_2$. Flows
 of rank 0 ($a^2=B_2/B_1$ and $a^2=-\tB_2/\tB_1$) constitute a subset of
 these solutions.

 The flows of classes $B$ and $C$ studied in detail in Papers I and II,
 are the subsets of the rank 1 flows with $B_1\ne 0,\ B_2=0$ and
 $B_1=0,\ B_2\ne 0$ respectively. Let us consider now the case with both
 nonzero $B_1$ and $B_2$. We denote $a_*^2=|B_2/B_1|$, $\xi=-B_1x$ and
 normalize $a$, $c$, and $U$ by $a_*>0$. It should be
 emphasized that both here and in Section~4, the variable $\xi$ is (up to
 sign) the $x$ coordinate scaled by the characteristic scale of flow
 variation.

 If the signs of $B_1$ and $B_2$ are the same, Eq.~(\ref {Eq-a1}) takes
 the form
 \be
 \dfrac{\dd a(\xi)}{\dd\xi} = \dfrac{a^2(\xi)-1}{c^2(\xi)-U^2(\xi)}\,c(\xi)
 \equiv \dfrac{a^2(\xi)-1}{c^4(\xi)-a^4(\xi)}\,c^3(\xi),
 \label{a1-}
 \ee
 and if they are opposite,
 \be
 \dfrac{\dd a(\xi)}{\dd\xi} = \dfrac{a^2(\xi)+1}{c^2(\xi)-U^2(\xi)}\,c(\xi)
 \equiv \dfrac{a^2(\xi)+1}{c^4(\xi)-a^4(\xi)}\,c^3(\xi).
 \label{a1+}
 \ee

 Let us first reveal characteristic features of the flows they describe.
 Equation (\ref{a1-}) has an obvious solution $a(\xi)=1$
 representing the set of rank 0 flows normalized by $a_*$. Imposing a
 small perturbation, $a(\xi)=1+\al(\xi)$, we see that
 \be
 \al(\xi) =
 \al_0\exp\left[2\IN{\xi}\!\dfrac{\dd yc^3(y)}{c^4(y)-1}\right]
 , \quad \al_0 = \mbox{const},
 \label{a=1}
 \ee
 tends to zero when $\xi\to-\infty$ in subcritical ($c(\xi)>1$) flows and
 when $\xi\to+\infty$ in supercritical ($c(\xi)<1$) ones.
 Thus, rank 0 flows define {\it one-sided asymptotics} for a certain subset
 of rank 1 flows. To reveal other features, let us consider solutions for
 $H(\xi)=\mbox{const}$, bearing in mind that they provide qualitatively
 correct description for flows with a limited variation in depth,
 $0<H_-<H(x)<H_+<\infty$, as well.
 \subsubsection{Flows of constant depth}
 \label{sec3-21}
 \hspace\parindent
 In this case, $c(\xi)=c_0=\mbox{const}$ and
 $a(\xi)=\Bl[c_0U(\xi)\Br]^{1/2}$. If $c_0\ne 1$, integrating Eq.~(\ref{a1-})
 yields
 \be
 a^3 + 3a + \frac{3}{2}\,(1-c_0^4)\ln\left|\dfrac{1-a}{1+a}\right| =
 3c_0^3(\xi_0-\xi), \quad \xi_0=\mbox{const}.
 \label{a1-c}
 \ee
 There are three singularities limiting the domains of flow existence and
 variation, $a(\xi_0)=0$, $a(\xi_c)=c_0$, and $a=1$. In addition, the
 behavior of $a(\xi)$ differs significantly for $c_0<1$ and $c_0>1$ (see
 Figures~\ref{fg3}\,($a,\,b$)).
%
 \begin{figure}[!h]
 \epsfxsize=100mm
 \centerline{\epsfbox{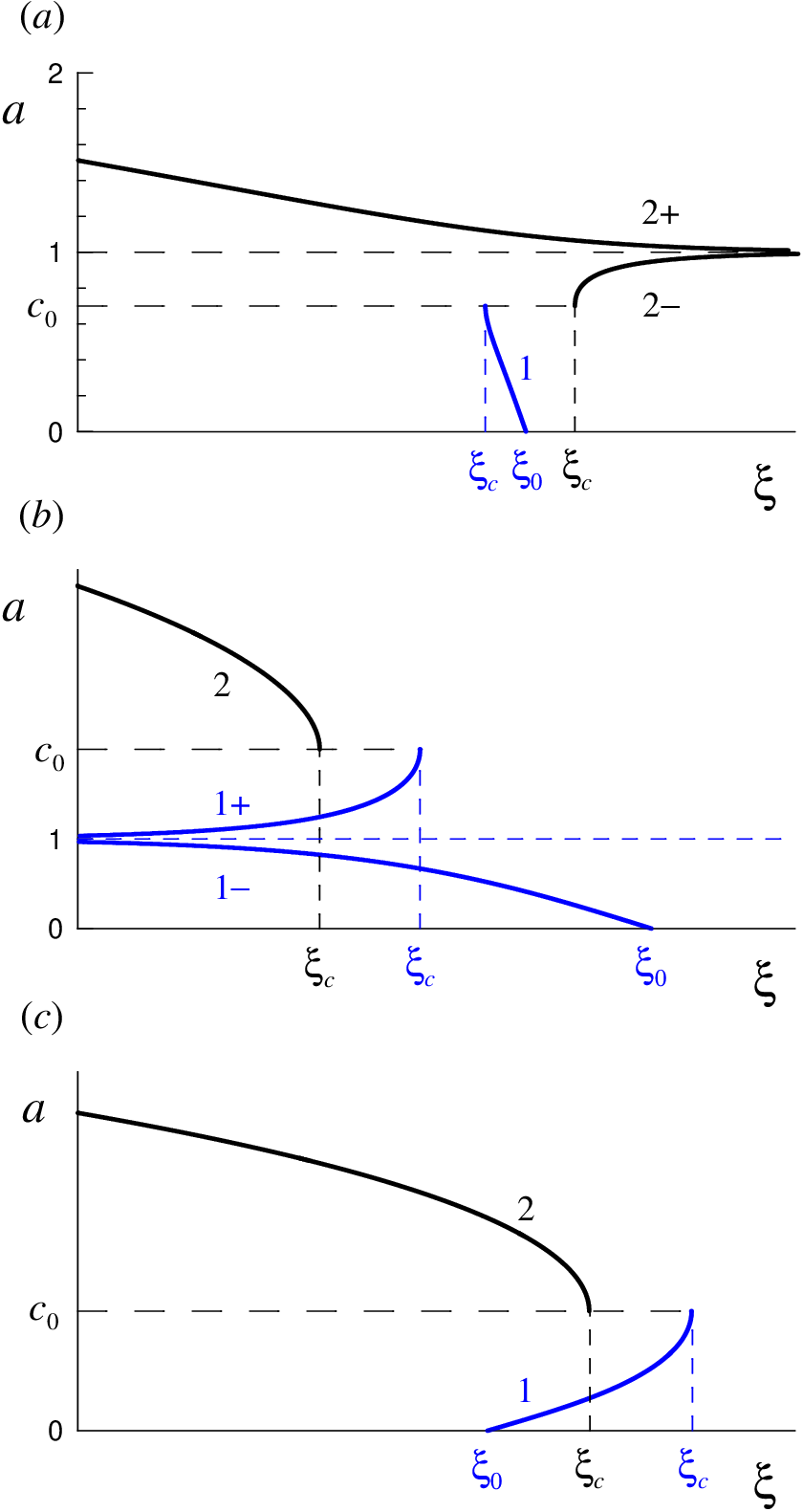}}
 \vspace{.2cm}
 \caption{Solutions to Eq.~(\ref{a1-c}) for\ \ ($a$)\ \ $c_0=0.7$
 and\ \ ($b$)\ \ $c_0=2$;\ \ ($c$)\ \ solution to Eq.~(\ref{a1+1})
 for $c_0=2$. Subcritical branches are marked by 1, supercritical ones --
 by 2. Signs `+' and `$-$' mark upper and lower sub-branches.}
 \label{fg3}
 \end{figure}

 Near the critical point
 \[
 \xi = \xi_c = \xi_0 - \frac{1}{3} - \dfrac{1}{c_0^2} -
 \dfrac{1-c_0^4}{3c_0^3}\,\ln\left|\dfrac{1-c_0}{1+c_0}\right|\,,
 \]
 the solution has a singularity of the fold type \cite{Arnold},
 \[
 a(\xi) = c_0 \pm \left[\dfrac{1-c_0^2}{2}(\xi-\xi_c)\right]^{1/2} -
 \dfrac{3+c_0^2}{12c_0}(\xi-\xi_c) + O\Bl(|\xi-\xi_c|^{3/2}\Br),
 \]
 and is defined only on one side of $\xi_c$, for $\xi\ge\xi_c$ when $c_0<1$
 and for $\xi\le\xi_c$ when $c_0>1$. Since $U=c_0\ne 0$ at the critical
 point, the flow continues on the other side of $\xi_c$, but ceases to be
 non-reflecting. To emphasize this fact, in Figure~\ref{fg3} and subsequent
 figures, the critical points of the subcritical and supercritical branches
 of the solutions are separated.

 The dashed lines $a=1$ are approached by solutions asymptotically, from
 below or above, in line with Eq.~(\ref{a=1}). Subcritical
 solutions are bounded from one or both sides. The supercritical solutions
 with $a>\max(c_0,\,1)$ grow indefinitely when $\xi\to-\infty$,
 \be
 a(\xi) = c_0(-3\xi)^{1/3} + O\Bl(|\xi|^{-1/3}\Br), \quad
 U(\xi) \sim(-\xi)^{2/3}.
 \label{as-a}
 \ee
 The class $B$ flows studied in Paper I can be obtained in the limit
 $c_0\gg 1$. Graphically, this corresponds to the merging of the $\xi$ axis
 with the $a=1$ line and the disappearance of the $1-$ branch of the
 solution in Figure~\ref{fg3}\,(b) (cf. Figure~I.3\,(a)).

 If $c_0=1$, the critical point disappears, and equation (\ref{a1-}) yields
 \be
 a^3(\xi) + 3a(\xi) = 3(\xi_0-\xi), \quad \xi_0 = \mbox{const}.
 \label{a1-1}
 \ee
 The flow described by this equation is defined for $\xi<\xi_0$ and singular
 at $\xi=\xi_0$: $U(\xi_0)=0,\ D\to\infty$. As $\xi$ decreases, $a$ and $U$
 increase monotonically in accordance with Eq.~(\ref{as-a}).
 The graph of $a(\xi)$ can be obtained from Figure~\ref{fg3}\,(b)
 in the limit $c_0\to 1$, when the $1+$ branch disappears, whereas the $1-$
 and 2 branches merge into a single continuous curve extending from
 $\xi=-\infty$ to $\xi=\xi_0$.

 Integrating Eq.~(\ref{a1+}) yields
 \be
 a^3 - 3a - 3(c_0^4-1)\arctan a = 3c_0^3(\xi_0-\xi), \quad
 \xi_0=\mbox{const}.
 \label{a1+1}
 \ee
 Here, the flow velocity vanishes at $\xi=\xi_0$ as well. The left-hand side
 of Eq.~(\ref{a1+1}) has a minimum at $a=c_0$, so at the critical point
 \[
 \xi = \xi_c = \xi_0 - \frac{1}{3} + \dfrac{1}{c_0^2} +
 \dfrac{c_0^4-1}{c_0^3}\,\arctan c_0 > \xi_0
 \]
 there is a fold-type singularity, and the solution is defined only for
 $\xi\le\xi_c$ (Figure~\ref{fg3}\,($c$)). In the neighborhood of $\xi_c$
 \[
 a(\xi) = c_0 \pm \left[\dfrac{1+c_0^2}{2}(\xi_c-\xi)\right]^{1/2} -
 \dfrac{3-c_0^2}{12c_0}(\xi_c-\xi) + O\Bl[(\xi_c-\xi)^{3/2}\Br],
 \]
 and for $\xi\to-\infty$\ \ $a(\xi)\approx c_0(-3\xi)^{1/3}$ (cf.
 (\ref{as-a})).
 The class $C$ flows studied in Paper II behave qualitatively
 in the the same manner (cf. Figure~II.2).

 Thus, in the case of constant-depth flows, equation (\ref{a1-}) has a global
 solution defined on the entire $\xi$ axis only for $c_0<1,\ a>1$.
 In all other cases, and in the case of equation (\ref{a1+}), the domain of
 the flow existence is limited by either the point where its velocity
 vanishes, or the critical point, or both. Let us now search for conditions
 under which the flow can be global.
 \subsubsection{Global solutions of equation (\ref{a1-})}
 \label{sec3-22}
 \hspace\parindent
 For $c(\xi)$ given, each particular solution of Eq.~(\ref{a1-}) or (\ref{a1+})
 can conveniently be represented on the half-plane $(\xi,\,a)$ as a
 trajectory $a(\xi)$, global or bounded. To find conditions for the existence
 of global solutions, we introduce the variables
 \be
 F(\xi) = \left[\dfrac{U(\xi)}{c(\xi)}\right]^{1/2}, \quad
 f(\xi) = \dfrac{1}{F(\xi)}\,,
 \label{Ff}
 \ee
 which are local Froude number and inverse Froude number, denote
 \be
 I_{F\pm}(\xi) = \pm\din_{\xi}^{\pm\!\infty}\dfrac{\dd y}{c(y)}\,, \quad
 I_{f\pm}(\xi) = \pm\din_{\xi}^{\pm\!\infty}c^3(y)\,\dd y,
 \label{IFf}
 \ee
 and rewrite Eq.~(\ref{a1-}) in different equivalent forms:
 \be
 \dfrac{\dd}{\dd\xi}\,\ln\left|\dfrac{a-1}{a+1}\right| = \dfrac{2}{c(1-F^4)}\,,
 \qquad \dfrac{\dd F}{\dd\xi} = \dfrac{F^2-c^{-2}}{1-F^4} -
 \dfrac{F}{c}\,\dfrac{\dd c}{\dd\xi}\,,
 \label{a1-F}
 \ee
 and
 \be
 \dfrac{\dd}{\dd\xi}\left(\dfrac{a^3}{3} + a + \frac{1}{2}\,
 \ln\left|\dfrac{a-1}{a+1}\right|\right) = -\dfrac{c^3}{1-f^4}\,, \qquad
 \dfrac{\dd f}{\dd\xi} = \dfrac{f^4-c^{-2}f^6}{1-f^4} +
 \dfrac{f}{c}\,\dfrac{\dd c}{\dd\xi}\,.
 \label{a1-f}
 \ee

 {\bf Subcritical flows ($F<1$).} Let $c(\xi)\le c_M<1$ everywhere. Then
 $a(\xi)$ decreases monotonically and reaches singular points $F=1$ and $F=0$
 at finite $\xi$, just as shown in Figure~\ref{fg3}$(a)$, and there are no
 global flows. Indeed, according to the first Eq.~(\ref{a1-F}),
 \[
 \dfrac{\dd}{\dd\xi}\,\ln\dfrac{1-a}{1+a} = \dfrac{2}{c(1-F^4)} > 2,
 \]
 hence, the transition from $a=0$ to $a=c<1$ takes a finite change in $\xi$.

 Now let $c(\xi)\ge c_m>1$ everywhere. Consider a trajectory passing through
 a point ($\xi_1,\,a_1$). As $\xi\to-\infty$,\ \ $a(\xi)\to 1$, from above
 ($a_1>1$) or from below ($a_1<1$). Integrating the first Eq.~(\ref{a1-F}),
 \be
 \ba{l}
 \ln\dfrac{|1-a(\xi)|}{1+a(\xi)} = \ln\dfrac{|1-a_1|}{1+a_1} +
 2\din_{\xi_{1}}^{\xi}\dfrac{\dd y}{c(y)[1-F^4(y)]}\ \
 \stackrel{\xi\to\infty}{\longrightarrow} \\ \\
 \ln\dfrac{1-a_1}{1+a_1}+
 \dfrac{2}{1-F^4(\xi_{1a})}\,I_{F+}(\xi_1), \quad \xi_{1a}>\xi_1,
 \ea
 \label{aIF}
 \ee
 we see that global trajectories can exist if $c(\xi)$ increases as
 $\xi\to+\infty$ fast enough for the integral $I_{F+}(\xi_1)$ to converge.
 Indeed, if $a(\xi)$ is bounded above, we can choose $a_1$ such that the
 right-hand side of Eq.~(\ref{aIF}) is negative. Then $a(\xi)\to a_0>0$ as
 $\xi\to+\infty$, and the flow velocity and channel width tend to zero,
 $D(\xi)\sim U(\xi)\approx a_0^2/c(\xi)$. In the case $a_1<1$, the set of
 such trajectories is bounded below by a global trajectory with $a_0=+0$, on
 which, in the case $c(\xi)\approx c_1\xi^\beta$, $\beta>1$, the flow
 velocity decreases as well while the channel width can even increase,
 \be
 a \sim \xi^{1-\beta}, \quad U = a^2/c \sim \xi^{2-3\beta}, \quad
 D \sim \xi^{\beta-2}, \quad H \sim c^2 \sim \xi^{2\beta}.
 \label{as-sub-}
 \ee

 If $a(\xi)$ grows with no limit as $\xi\to+\infty$, then the left-hand side
 of Eq.~(\ref{aIF}) tends to zero and $F(\xi)<1$ everywhere. Thus, in the
 case $c(\xi)\approx c_1\xi^\beta$,
 \[
 a \sim \xi^{\beta-1}, \quad F \sim \xi^{-1}, \quad U \sim \xi^{\beta-2},
 \quad D \sim \xi^{2-3\beta}.
 \]

 {\bf Supercritical flows.} Here, it is convenient to use the variable $f$
 and Eqs.~(\ref{a1-f}). If $c<1$ everywhere, all trajectories with
 $a_1>1$ are global (see Figure~\ref{fg3}($a$), the upper curve). In the
 domain $c(\xi)<a(\xi)<1$,\ \ $a(\xi)$ increases monotonically and tends
 to 1 as $\xi\to+\infty$. Integrating the first Eq.~(\ref{a1-f})
 with Eq.~(\ref{IFf}) taken into account,
 \[
 \ba{l}
 \dfrac{a^3(\xi)}{3} + a(\xi) + \frac{1}{2}\,
 \ln\dfrac{|1-a(\xi)|}{1+a(\xi)} =
 \dfrac{a_1^3}{3} + a_1 + \frac{1}{2}\,\ln\dfrac{|1-a_1|}{1+a_1}
 -\!\din_{\xi_1}^{\xi}\!\dfrac{\dd\xi_1c^3}{1-f^4(\xi_1)}
    \\ \\
 \stackrel{\xi\to-\infty}{\longrightarrow}\ \
 \dfrac{a_1^3}{3} + a_1 + \frac{1}{2}\,\ln\dfrac{|1-a_1|}{1+a_1} +
 \dfrac{I_{f-}(\xi_1)}{1-f(\xi_b)}\,, \quad \xi_b < \xi_1,
 \ea
 \]
 shows that for global trajectories to exist, the convergence of
 $I_{f-}(\xi_1)$ is sufficient.

 Finally, if $c(\xi)>1$ for $\xi<\xi_-$, then, on a supercritical trajectory,
 $a(\xi)$ decreases and inevitably reaches $f=1$ at finite $\xi$, unless
 $c(\xi)$ becomes less than 1 for $\xi>\xi_+$. Then there are global
 trajectories on which $a(\xi)\to 1+0$ when $\xi\to+\infty$. When $c(\xi)>1$
 for $\xi>\xi_+$, there are no global trajectories.
 \subsubsection{Global solutions of equation (\ref{a1+})}
 \label{sec3-23}
 \hspace\parindent
 Using notations (\ref{Ff}), we write equation Eq.~(\ref{a1+}) in different
 equivalent forms:
 \bea
 \label{Eq-a+}
 \dfrac{\dd}{\dd\xi}\,\arctan a & = & \dfrac{1}{c(1-F^4)}\,, \\
 \label{Eq-F+}
 \dfrac{\dd F}{\dd\xi} & = & \dfrac{F^2+c^{-2}}{1-F^4} - M\,F, \quad
 M(\xi) = \dfrac{\dd\ln c}{\dd\xi}\,, \\
 \label{Eq-f+}
 \dfrac{\dd}{\dd\xi}\left(\dfrac{a^3}{3} - a + \arctan a\right) & = &
 - \dfrac{c^3}{1-f^4}\,.
 \eea

 {\bf Subcritical flows ($F<1$).} Consider a trajectory passing through the
 point $(\xi_1,\,F_1)$. Along this trajectory, $a(\xi)$ increases
 monotonically. To be global, the trajectory must not end at either $F=0$
 when $\xi<\xi_1$ or $F=1$ when $\xi>\xi_1$. From Eq.~(\ref{Eq-a+}) and the
 mean-value theorem, we obtain ($\xi_{1b}$ lies between $\xi_1$ and $\xi$)
 \be
 \arctan a(\xi) - \arctan a(\xi_1) = \din_{\xi_1}^{\xi}\dfrac{\dd y}
 {[1-F^4(y)]c(y)} = \dfrac{1}{1-F^4(\xi_{1b})}\din_{\xi_1}^{\xi}
 \dfrac{\dd y}{c(y)}
 \label{a+I}
 \ee
 and we see that for $a(\xi)$ to be positive for $\xi<\xi_1$, we need such a
 growth of $c(\xi)$ as $\xi\to-\infty$ that $I_{F-}(\xi_1)$ not only
 converges, but is also sufficiently small, at least $I_{F-}(\xi_1)<\pi/2$.
 In this case, $F(\xi)$ increases and the sufficient condition for the
 existence of global trajectories is
 \be
 c(\xi_1) > a(\xi_1) \ge \tan\dfrac{I_{F-}(\xi_1)}{1-F_1^4}\,, \quad
 \dfrac{I_{F-}(\xi_1)}{1-F_1^4} < \dfrac{\pi}{2}\,.
 \label{glob1}
 \ee
 If the trajectory is not bounded on the left, then for $\xi\to-\infty$
 either $a\to a_0>0$ and $U\sim D\sim c^{-1}$, or $a\to +0$. In the latter
 case, if $c(\xi)\sim(-\xi)^\beta,\ \beta>1$ (cf. Eq.~(\ref{as-sub-})),
 \[
 a \sim (-\xi)^{1-\beta}, \quad F \sim (-\xi)^{1-2\beta}, \quad
 U \sim (-\xi)^{2-3\beta}, \quad D \sim (-\xi)^{\beta-2}, \quad
 H \sim (-\xi)^{2\beta}.
 \]

 Similar reasoning shows that for the unlimited continuation of a trajectory
 into the $\xi>\xi_1$ domain, the convergence of $I_{F+}(\xi_1)$ is needed
 together with fulfillment of the inequality
 \[
 \arctan a(\xi_1) + \dfrac{I_{F+}(\xi_1)}{1-F^4(\xi_{1b})} <  \dfrac{\pi}{2}\,.
 \]
 Unlike Eq.~(\ref{glob1}), this is facilitated by a decrease rather than
 increase in $a(\xi_1)$. To reconcile these conditions, we consider the
 null-isocline (NI) of Eq.~(\ref{Eq-F+}). It is described by the equation
 \be
 F_N(1-F_N^4) = \dfrac{1}{M}\left(F_N^2 +\dfrac{1}{c^2}\right), \quad
 M(\xi) =\dfrac{\dd\ln c}{\dd\xi}\,,
 \label{F*}
 \ee
 the solution of which can be conveniently represented graphically (see
 Figure~\ref{fg4}).
%
 \begin{figure}[t]
 \epsfysize=60mm
 \centerline{\epsfbox{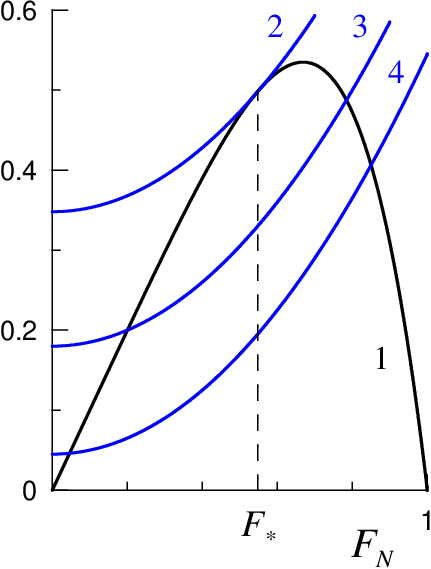}}
 \vspace{.2cm}
 \caption{Solution of Eq.~(\ref{F*}):\ \ line 1 shows the left-hand
 side, and lines 2, 3, and 4 show the right-hand side for increasing $c$;
 $M=2$.}
 \label{fg4}
 \end{figure}
 The left-hand side reaches its maximum value of $5^{5/4}/4$ at $F_N=5^{-1/4}$,
 hence, solutions can only exist if
 \[
 c^2(\xi)M(\xi) > \dfrac{5^{5/4}}{4},\ \ \mbox{or}\ \
 \dfrac{\dd c^2}{\dd\xi} \equiv 2c^2M > \dfrac{5^{5/4}}{2} \approx 3.7384.
 \]
 Figure~\ref{fg4} shows that the equation has two solutions, which merge and
 disappear (become complex-valued) as the parameters change. This occurs
 when the derivatives of the right-hand and left-hand sides of Eq.~(\ref{F*})
 are equal, at the point $F_N=F_*$ where $1-5\,F_*^4=2F_*/M$. Solutions
 exist if at $F_N=F_*$ the left-hand side of Eq.~(\ref{F*}) is not less than
 the right-hand side. Thus, we obtain the conditions for solution existence
 in a parametric form:
 \be
 \ba{l}
 M = \dfrac{2F_*}{1-5F_*^4}\,, \quad
 c^2 \ge c_*^2 = \dfrac{1-5F_*^4}{F_*^2(1+3F_*^4)}\,,
   \\ \\
 \dfrac{\dd c^2}{\dd\xi} = 2c^2M \ge \dfrac{4}{F_*(1+3F_*^4)}\,,
 \quad 0 < F_* < 5^{-1/4}.
 \ea
 \label{n-eq}
 \ee
 NI is located to the right of some point $\xi=\xi_*$ at which its upper and
 lower branches, $F=F_{N\pm}(\xi)$, meet (curve 1 in Figure~\ref{fg5}($a$)).
 Between the branches, $\dd F/\dd\xi<0$.
%
 \begin{figure}[t]
 \epsfysize=145mm
 \centerline{\epsfbox{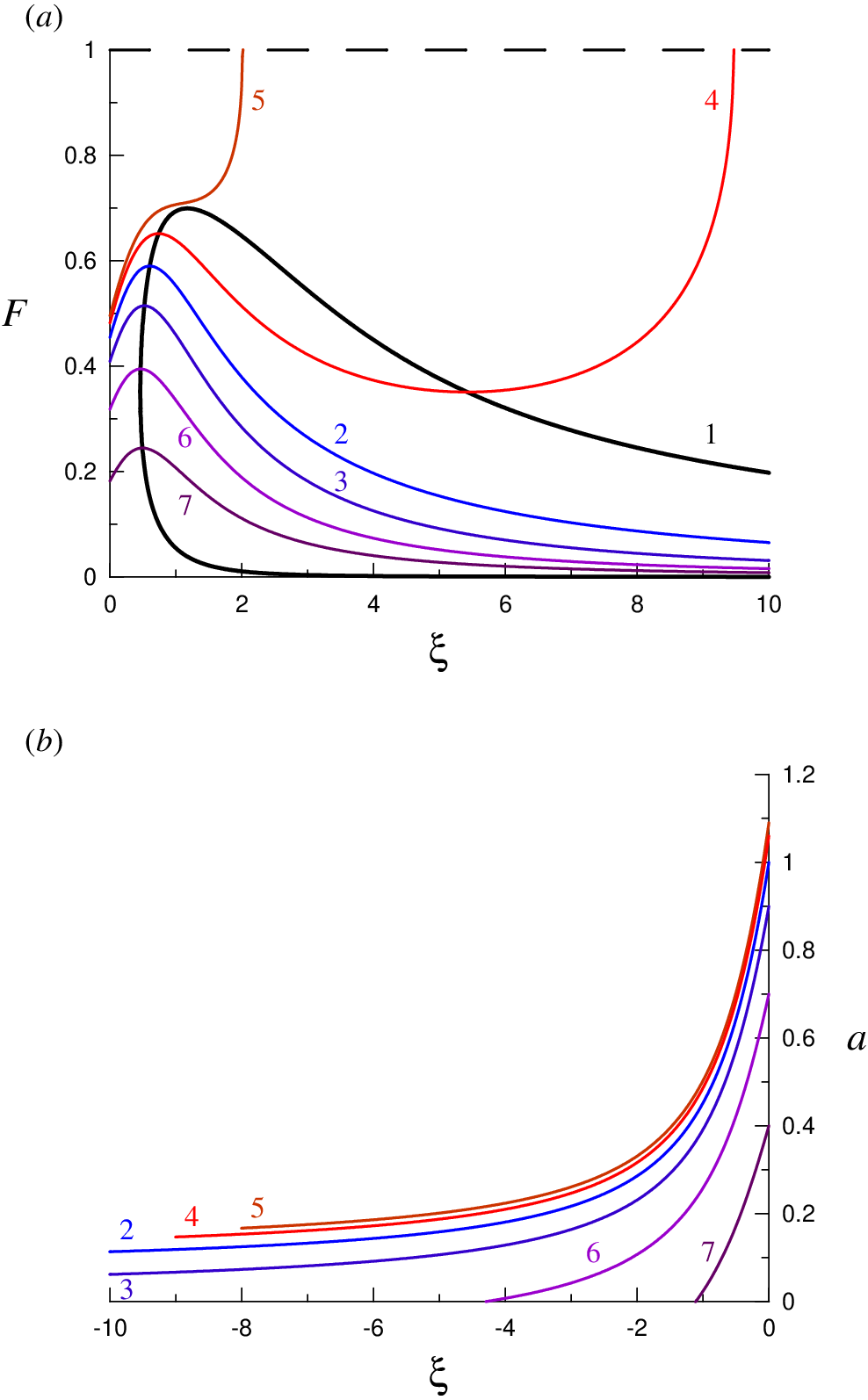}}
 \vspace{.2cm}
 \caption{The subcritical part of the phase portrait of Eq.~(\ref{a1+}):
 \ ($a$) -- $F(\xi)$ for $\xi\ge 0$,\ \ ($b$) -- $a(\xi)$ for $\xi\le 0$.
 Curve 1 corresponds to NI. Trajectories:\ \ 2\ \ $a(0)=1$,\ \ 3\ \ $a(0)=0.9$,\
 \ 4\ \ $a(0)=1.06$,\ \ 5\ \ $a(0)=1.09$,\ \ 6\ \ $a(0)=0.7$,\ \ 7\ \
 $a(0)=0.4$.}
 \label{fg5}
 \end{figure}

 Let us trace the evolution of $F(\xi)$ along a trajectory that approaches
 the NI from the left and intersects it. At the intersection point, $F(\xi)$
 has a local maximum, and the trajectory is unbounded on the right if it
 remains between the NI branches. Otherwise, it ends at the point $F=1$
 (like curve 4 in Figure~\ref{fg5}($a$)). The result depends both on the
 behavior of $c(\xi)$ and on the ``height'' of the trajectory's first
 intersection with NI.

 Figure~\ref{fg4} corresponds to $M(\xi)=\mbox{const}$ when $c(\xi)$ and
 the flow depth grow exponentially. In this case, $F_{N+}(\xi)\to F_{N0}-0$,
 $F_{N0}=\mbox{const}<1$, and $F_{N-}(\xi)\to 0$, and after crossing NI, any
 trajectory remains between its branches and continues to the right indefinitely.
 If $c(\xi)$ grows more slowly, for example, $c(\xi)\approx c_1\xi^\beta$,
 $\beta>1$, both $F_{N\pm}(\xi)\to 0$ as $\xi\to\infty$. As long as the
 trajectory lies between the branches, the equations for $F$ and $F_N$ can
 be approximately written as
 \[
 \dfrac{\dd F}{\dd\xi} = F^2 +\dfrac{1}{c_1^2\xi^{2\beta}} -
 \dfrac{\beta F}{\xi}\,, \qquad
 F_{N+} \approx \dfrac{\beta}{\xi} - \dfrac{1}{\beta c_1^2\xi^{2\beta-1}}\,,
 \quad F_{N-} \approx \dfrac{1}{\beta c_1^2\xi^{2\beta-1}}\,.
 \]
 Between the NI branches, $\dd F/\dd\xi$ has a minimum at $F=F_m=\beta/(2\xi)$,
 \[
 \left.\dfrac{\dd F}{\dd\xi}\right|_{F_m} = -\dfrac{\beta^2}{4\xi^2} +
 \dfrac{1}{c_1^2\xi^{2\beta}}\,,
 \]
 which for $\beta\ge 2$ is no greater than $\dd F_m/\dd\xi=-\beta/(2\xi^2)$.
 Therefore, if $F(\xi_s)=F_m(\xi_s)$ for some $\xi=\xi_s$, then
 $F(\xi)\le F_m(\xi)$ for $\xi>\xi_s$, and such a trajectory is not bounded
 on the right. The underlying trajectories are unbounded as well.

 Figure~\ref{fg5} shows a phase portrait of subcritical flows calculated for
 $c(\xi)=2.2\,(\xi^2+1)$. The trajectories shown in the figure differ in the
 value of $a(\xi=0)$ and represent all possible behavior types. The rapid
 increase in $c(\xi)$ as $\xi\to\pm\infty$ excludes trajectories bounded on
 both sides. Trajectories 2 and 3 are unbounded and belong to the bundle of
 global trajectories, trajectories 4 and 5 are bounded on the right by $F=1$,
 and trajectories 6 and 7 are bounded on the left by the point where the flow
 velocity vanishes. On the trajectories 2 through 5 $a(\xi)$ has a positive
 limit when $\xi\to-\infty$.

 {\bf Supercritical flows ($f<1$).} In this domain, $a(\xi)=c(\xi)/f(\xi)$
 decreases monotonically in accordance with the second Eq.~(\ref{Eq-a+}).
 Its integration
 \[
 \dfrac{a^3(\xi)}{3} - a(\xi) + \arctan a(\xi) = \dfrac{a_1^3}{3} - a_1 +
 \arctan a_1 - \!\din_{\xi_1}^{\xi}\!\dfrac{\dd y\,c^3(y)}{1-f^4(y)}
 \]
 shows that global trajectories exist if $I_{f+}(\xi_1)$ converges. When
 $\xi\to+\infty$, on such a trajectory either $a(\xi)\to a_0>0$ and the flow
 velocity and channel width grow indefinitely, $U\sim D\sim 1/c$, or
 $a(\xi)\to +0$. Then, if $c(\xi)\approx c_2\xi^{-\beta}$, $\beta>1/3$,
 \[
 a \sim \xi^{(1-3\beta)/5} \to 0, \quad f \sim \xi^{-(1+2\beta)/5} \to 0,
 \quad  D \sim \xi^{(11\beta-2)/5} \to \infty, \quad U \sim \xi^{(2-\beta)/5}.
 \]
 \section{Currents of rank 2}
 \label{sec4}
 \hspace\parindent
 For accelerated waves, Eqs.~(\ref{Eq+}) and (\ref{Eq+1}) for $N=2$ yield
 \[
 \ba{l}
 V_2 = \dfrac{a}{U+c}\,, \quad W_2 = 0, \quad W_1 = -\dfrac{U+c}{c}\,
 \dfrac{\dd a}{\dd x}\,,
   \\ \\
 \dfrac{\dd}{\dd x}\left[(U+c)\left(V_1+\dfrac{\dd a}{\dd x}\right)\right] =
 \dfrac{B_1a^2-B_2}{U^2-c^2}\,c,
   \\ \\
 \dfrac{\dd}{\dd x}\left[(U+c)\left(V_1+\dfrac{U}{c}\,\dfrac{\dd a}{\dd x}
 \right)\right] - \dfrac{2}{a}\,\dfrac{\dd a}{\dd x}\left[(U+c)\left(V_1+
 \dfrac{U}{c}\,\dfrac{\dd a}{\dd x}\right)\right] =
 -\dfrac{B_1a^2-B_2}{U^2-c^2}\,c.
 \ea
 \]
 Adding and subtracting the last two equations, we get
 \[
 \ba{l}
 \dfrac{\dd}{\dd x}\Bl[(U+c)V_1\Br] - \dfrac{1}{a}\,\dfrac{\dd a}{\dd x}
 (U+c)V_1 = \dfrac{U(U+c)}{a\,c}\left(\dfrac{\dd a}{\dd x}\right)^2 -
 \dfrac{\dd}{\dd x}\left[\dfrac{(U+c)^2}{2c}\,\dfrac{\dd a}{\dd x}\right],
    \\ \\
 \dfrac{1}{a}\,\dfrac{\dd a}{\dd x}(U+c)V_1 =
 \dfrac{\dd}{\dd x}\left(\dfrac{U^2-c^2}{2c}\,\dfrac{\dd a}{\dd x}\right) -
 \dfrac{U(U+c)}{a\,c}\left(\dfrac{\dd a}{\dd x}\right)^2 +
 \dfrac{B_1a^2-B_2}{U^2-c^2}\,c.
 \ea
 \]
 Eliminating $V_1$, we arrive at an equation relating $U(x)$ and $c(x)$ in
 currents of rank 2:
 \be
 \ba{l}
 \dfrac{\dd}{\dd x}\left[\dfrac{a}{\dd a/\dd x}\,\dfrac{\dd}{\dd x}\left(
 \dfrac{U^2\!-\!c^2}{c}\,\dfrac{\dd a}{\dd x}\right)\right] -
 2\dfrac{\dd}{\dd x}\left(\dfrac{U^2\!-\!c^2}{c}\,\dfrac{\dd a}{\dd x} -
 \dfrac{B_1a^2-B_2}{U^2\!-\!c^2}\,\dfrac{a\,c}{\dd a/\dd x}\right)
   \\ \\ \phantom{wwwwwwwwwwwwwwwwwn}
 - 2\,\dfrac{B_1a^2-B_2}{U^2-c^2}\,c = 0.
 \ea
 \label{R2}
 \ee
 For further consideration, it is convenient to introduce the variable $s$,
 \be
 \dd s = \dfrac{c(x)\,\dd x}{U^2(x)-c^2(x)}\ \ \ \Longrightarrow\ \ \
 \dfrac{U^2-c^2}{c}\,\dfrac{\dd}{\dd x} = \dfrac{\dd}{\dd s}\,,
 \label{sx}
 \ee
 consider $a$, $c$, and $U$ as $s$-dependent, and denote the $s$-derivative
 by prime, $(\dots)'=\dd(\dots)/\dd s$. Multiplying Eq.~(\ref{R2}) by
 $(U^2-c^2)/c$, after simple algebra we obtain the equation
 \be
 \left(\dfrac{a\,a''}{a'}\right)' - 2a'' + 2\left[\dfrac{a}{a'}\,
 (B_1a^2-B_2)\right]' - 2(B_1a^2-B_2) =0,
 \label{Eq-a2-1}
 \ee
 as well as equivalent equations
 \be
 \left(\dfrac{a''}{a\,a'}\right)' + 2\left(\dfrac{B_1a^2-B_2}{a\,a'}
 \right)' + 2\dfrac{B_1a^2-B_2}{a^2} =0
 \label{Eq-a2-2}
 \ee
 and
 \be
 a\,a'a''' - a\,{a''}^2 - {a'}^2a'' + 4B_1a^2{a'}^2 -
 2(B_1a^2-B_2)a\,a'' = 0.
 \label{Eq-a2-3}
 \ee
 In addition, we shall need the equations following from Eqs.~(\ref{Eq-a2-1})
 -- (\ref{Eq-a2-3}) for $b(s)=1/a(s)$:
 \be
 \ba{l}
 \left(\dfrac{b''}{b\,b'}\right)' + 2\left(\dfrac{B_2b^2-B_1}{b\,b'}
 \right)' + 2\dfrac{B_2b^2-B_1}{b^2} =0,
   \\ \\
 b\,b'b''' - b\,{b''}^2 - {b'}^2b'' + 4B_2b^2{b'}^2 -
 2(B_2b^2-B_1)b\,b'' = 0.
 \ea
 \label{Eq-b2}
 \ee
 For retarded waves with $n=2$, similar calculations (or the substitution
 indicated after Eq.~(\ref{Eq-1})) lead to the equation
 \[
 a\,a'a''' - a\,{a''}^2 - {a'}^2a'' - 4\tB_1a^2{a'}^2 +
 2(\tB_1a^2+\tB_2)a\,a'' = 0,
 \]
 which, due to the arbitrariness of the constants $\tB_{1,2}$, describes the
 same set of flows as Eq.~(\ref{Eq-a2-3}).

 As expected, flows of rank 2 include the flows of ranks 0 and 1. Indeed,
 Eq.~(\ref{Eq-a2-3}) has an obvious solution $a=\mbox{const}$. Moreover, it
 can be represented as
 \[
 \left(\dfrac{1}{a\,a'}\Bl[a'-(B_1a^2-B_2)\Br]'\right)' -
 \dfrac{2}{a}\left(\dfrac{a'-(B_1a^2-B_2)}{a'}\right)' = 0,
 \]
 and we see that it is satisfied by solutions of Eq.~(\ref{Eq-a1})
 which can be written as $a'=B_1a^2-B_2$.

 We face an extremely difficult problem of searching for solutions to a
 third-order nonlinear equation, in which the independent variable $s$ is
 related to the coordinate $x$ by the relation (\ref{sx}) depending on the
 velocities $U$ and $c$. Leaving this problem for later, we shall limit
 ourselves to studying simpler cases where one of the constants $B_{1,\,2}$
 is equal to zero. If $B_2=0$, we can integrate Eq.~(\ref{Eq-a2-2}),
 \be
 a'' + 2B_1(s-s_a)\,a'a + 2B_1a^2 = 0, \quad s_a=\mbox{const},
 \label{a2}
 \ee
 and if $B_1=0$, we can integrate the first equation (\ref{Eq-b2}),
 \be
 b'' + 2B_2(s-s_b)\,b'b + 2B_2b^2 = 0, \quad s_b=\mbox{const}.
 \label{b2}
 \ee

 These equations differ only in notation and are equivalent to the equation
 \be
 y'' + 2B(s-s_0)\,y'y + 2By^2 = 0,
 \label{Eq-y}
 \ee
 the solution of which can be written in parametric form (see
 Appendix~\ref{AppB})
 \be
 s-s_0 = \dfrac{\tau_*-\tau}{C}\,, \quad y(\tau) = \dfrac{C^2}{BE(\tau)}\,;
 \quad E(\tau) = (\tau-\tau_*)\coth\tau - 1, \ \ \tau_* = \mbox{const},
 \label{Sol}
 \ee
 where $\tau$ is a parameter, and the coefficient $C$ relating $s$ and $\tau$
 yet determines the scale of the $y(\tau)$ variation. The function $E(\tau)$
 has a single real zero at $\tau=\tau_0$ obeying the equation
 \[
 \tau_0 - \tanh\tau_0 = \tau_*, \quad \tau_0\tau_*\ge 0.
 \]
 The function $a(\tau)$ is proportional to either $E$ (Eq.~(\ref{b2}))
 or $1/E$ (Eq.~(\ref{a2})). Figure~\ref{fg6} shows graphs of both
 $E(\tau)$ (in black) and $1/E(\tau)$ (in blue) for positive values of
 $\tau_*$.
%
 \begin{figure}[!h]
 \epsfxsize=120mm
 \centerline{\epsfbox{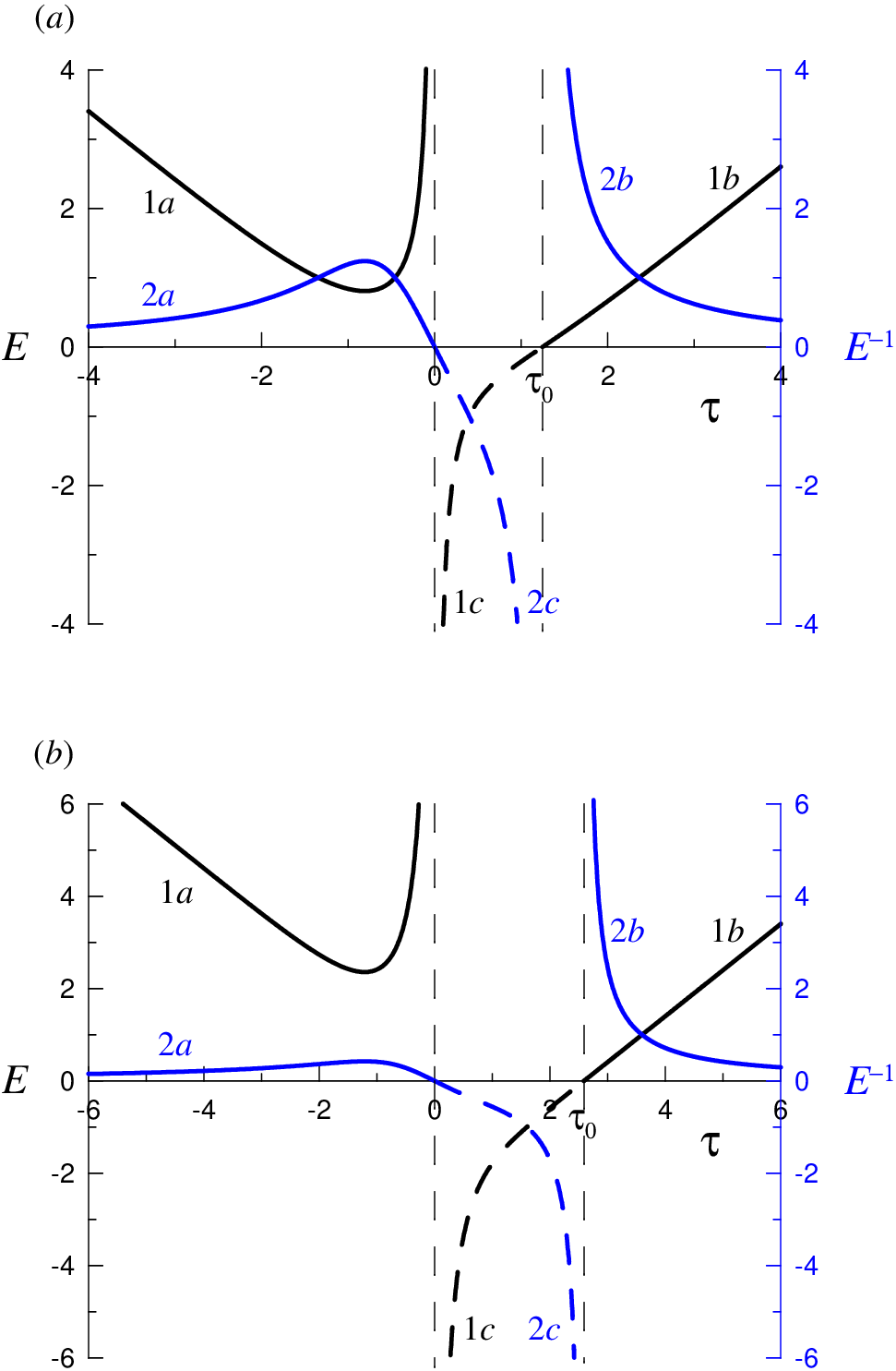}}
 \vspace{.2cm}
 \caption{The plots of $E(\tau)$ (black lines $1a,\,b,\,c$) and $1/E(\tau)$
 (blue lines $2a,\,b,\,c$) for ($a$) $\tau_*=0.4$ and ($b$) $\tau_*=1.6$.
 Negative branches are shown by dashes.}
 \label{fg6}
 \end{figure}
 Graphs for negative $\tau_*$ can be obtained by mirroring them with respect
 to the vertical line $\tau=0$. The asymptotic expansions of $E(\tau)$ are
 as follows (for definiteness, we assume $\tau_*>0$):
 \be
 \ba{ll}
 \tau \to -\infty: & E = -\tau\!+\!(\tau_*\!-\!1) + O\Bl(\tau\re^{2\tau}\Br),
   \\ \\
 \tau \to +\infty: & E = \tau - (\tau_* + 1) + O\Bl(\tau\re^{-2\tau}\Br);
   \\ \\
 \tau \to 0: & E = -\dfrac{\tau_*}{\tau}\left(1+\dfrac{\tau^2}{3}\right) +
 O(\tau^2),
   \\ \\
 \tau \to \tau_0: & E = (\tau_0\!-\!\tau_*)(\tau\!-\!\tau_0)\!+
 \!O\Bl[(\tau\!-\!\tau_0)^3\Br].
 \ea
 \label{E-as}
 \ee
 \subsection{Flows described by the solution (\ref{Sol})}
 \label{sec4-1}
 \hspace\parindent
 The condition $a>0$ ($y>0$) imposed by the problem statement confines
 the existence domain of the flow to one of three intervals into which
 the $\tau$-axis is divided according to the sign of $E$ (see
 Figure~\ref{fg6}). In the lateral intervals, $-\infty<\tau<0$ and
 $\tau_0<\tau<\infty$,\ \ $E(\tau)>0$, therefore, the flows corresponding to
 $B_{1,\,2}>0$ are defined here, and the middle interval $0<\tau<\tau_0$
 belongs to flows with $B_{1,\,2}<0$. Our goal is to study the qualitative
 behavior of non-reflecting flows. For convenience, we ``normalize'' the
 solutions of Eq.~(\ref{a2}) by setting $C=|B_1|^{1/2}$, and divide the
 flows they describe into classes $\hB_+$ and $\hB_-$ according to the sign
 of $B_1$. Similarly, for the solutions of Eq.~(\ref{b2}), we set
 $C=|B_2|^{1/2}$, and divide the flows into classes $\hC_+$ and $\hC_-$. Then,
 $a(\tau)=\pm 1/E(\tau)$ in classes $\hB_\pm$ (see curves $2a,\,b,\,c$ in
 Figure~\ref{fg6}) and $a(\tau)=\pm E(\tau)$ in classes $\hC_\pm$ (see curves
 $1a,\,b,\,c$).

 As can be seen from the figure, at the boundaries of each domain $a(\tau)$
 has singularities $a=0$ and/or $a\to\infty$. The fold-type singularity of
 $a(x)$ at the critical point $x_c$, where $a=U=c$ (see Figure~\ref{fg3}),
 inherent to flows of rank 1, arises because upon passing this point, the
 dependence of the coordinate $x$ on the parameter $\tau$ ceases to be
 monotonic (see the first formula (\ref{Sol}) and Eq.~(\ref{sx})),
 \be
 \dd\tau = -C\dd s = C\,\dfrac{c\,\dd x}{c^2-U^2} \equiv
 \dfrac{c\,\dd\xi}{c^2-U^2} \equiv \dfrac{c^3\dd\xi}{c^4-a^4}
 \equiv \dfrac{\dd\xi}{c(1-F^4)}\,, \qquad \xi = C\,x.
 \label{xtau}
 \ee
 It should be borne in mind that in subcritical $(a<c,\ F<1)$ flows the
 signs of $\dd\tau$ and $\dd\xi$ coincide, while in supercritical
 $(a>c,\ F>1)$ ones they are opposite. Now, using the variable $\xi$ instead
 of $x$, we proceed to analyze the behavior of the trajectories $a(\xi)$ in
 various cases assuming the $c(\xi)$ profile to be given and relying on
 Eq.~(\ref{xtau}) in the form
 \be
 \dfrac{\dd\xi}{c(\xi)} = \Bl(1-F^4\Br)\dd\tau \quad \mbox{or} \quad
 c^3(\xi)\dd\xi = \Bl(c^4 - a^4\Br)\dd\tau.
 \label{xtau1}
 \ee
 \subsubsection{Currents of constant depth ($c(\xi)=c_0=\mbox{const}$)}
 \label{sec4-11}
 \hspace\parindent
 The interval $0<\tau<\tau_0$ (see Figure~\ref{fg6}) is associated with
 flows of the classes $\hB_-$ (curve $2\,c$) and $\hC_-$ (curve $1\,c$).
 Here, all three singularities, $a=0$, $a=c_0$, and $a\to\infty$, correspond
 to finite $\tau$. But on the $\xi$ axis, the latter is reached only
 asymptotically, when $|\xi|\to\infty$. Indeed, $a(\tau)$ has a first-order
 pole at this point, the integral $\int a^4(\tau)\dd\tau$ diverges and,
 according to Eq.~(\ref{xtau}), $a(\xi)\sim |\xi|^{1/3}$, cf. Eq.~(\ref{as-a}).
 For flows of class $\hB_-$, the dependence $a(\xi)$ is similar to that shown
 in Figure~\ref{fg3}\,($c$), whereas $a(\xi)$  for flows of class $\hC_-$
 can be obtained by replacing $\xi$ with $-\xi$ in this figure.

 In the $\tau>\tau_0$ domain, $a(\xi)$ for flows of class $\hC_+$ (curve
 $1\,b$) is also similar to that shown in Figure~\ref{fg3}\,($c$). As for
 flows of class $\hB_+$ (curve $2\,b$), only the critical point corresponds
 to finite $\xi$, the other singularities are reached when $\xi\to+\infty$.
 Therefore, for these flows $a(\xi)$ is similar to that shown in
 Figure~\ref{fg9}($a$).

 A completely different type of behavior of $a(\xi)$ is inherent to flows
 associated with the $\tau<0$ domain. The singularities at its ends are
 identical, hence, $a(\tau)$ has a maximum $a_M(\tau_*)$ in flows of class
 $\hB_+$ (curves $2\,a$ in Figure~\ref{fg6}) and a minimum $a_m(\tau_*)$ in
 flows of class $\hC_+$ (curves $1\,a$). As a result, for $c_0>a_M$, the
 $\hB_+$ class flows are subcritical and bounded on the right by $a=0$ at
 $\xi_0=\xi(\tau=0)$ (Figure~\ref{fg7}\,$(a)$). On the other hand, if
 $c_0<a_m$, the $\hC_+$ class flows are supercritical and global
 (Figure~\ref{fg7}\,$(b)$). For clarity, these figures show
 $F(\xi)=a(\xi)/c_0$ and $f(\xi)=1/F(\xi)$ respectively.
%
 \begin{figure}[!h]
 \epsfysize=100mm
 \centerline{\epsfbox{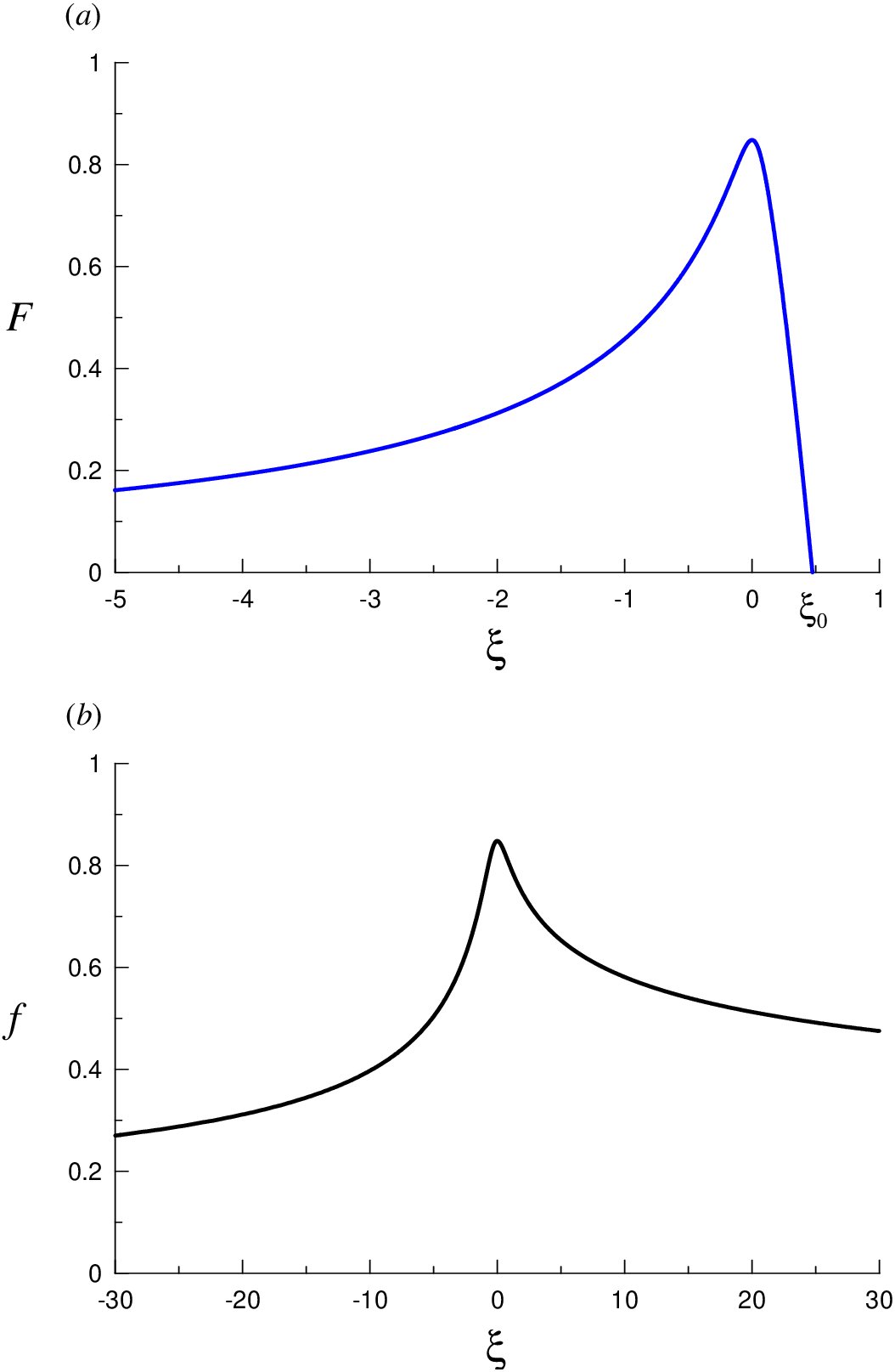}}
 \vspace{.2cm}
 \caption{Currents without critical points: $(a)$ -- flows of class $\hB_+$
 (subcritical), $c_0=0.5$, and $(b)$ -- flows of class $\hC_+$ (supercritical),
 $c_0=2$; $\tau_*=1.6$ for both.}
 \label{fg7}
 \end{figure}
 We call attention to the fact that $F(\xi)$ and $f(\xi)$ have significantly
 different rates of decreasing with distance from the maximum. This can be
 explained using the first Eq.~(\ref{xtau1}). Indeed, in $\hB_+$ flows,
 $F\to 0$ at both ends of the interval $-\infty<\tau<0$. Therefore, the
 $\xi(\tau)$ dependence resembles a linear one, and
 $F\sim -1/(c_0\tau)\sim -1/\xi$ as $\xi\to-\infty$ ($\tau\to-\infty$) (see
 formulas (\ref{E-as})). Conversely, in $\hC_+$ flows $F\to+\infty$ at both
 ends of the interval. Therefore, as $\tau\to 0$,
 \be
 F(\tau) \sim -\dfrac{\tau_*}{c_0\tau}\,, \quad
 \xi \sim \dfrac{\tau_*^4}{3c_0^3\tau^3}\ \to -\infty, \quad f(\xi) \sim
 \left(-\dfrac{3\xi}{\tau_*}\right)^{-1/3},
 \label{as1C+}
 \ee
 and for $\tau\to-\infty$
 \be
 F(\tau) \sim -\dfrac{\tau}{c_0}\,, \quad \xi \sim -\dfrac{\tau^5}{5c_0^3}\
 \to +\infty, \quad f(\xi) \sim \left(\dfrac{5\xi}{c_0^2}\right)^{-1/5}.
 \label{as2C+}
 \ee

 When $c_0<a_M$, the solutions describing flows of class $\hB_+$ have two
 critical points, at each of which $a(\xi)$ has a singularity of the fold
 type (Figure~\ref{fg8}\,$(a)$). In this case, there are subcritical
 (curve 1) and supercritical (curve 2) flows bounded on both sides, and a
 subcritical flow (curve 3) bounded on the right. Solutions describing flows
 of class $\hC_+$ for $c_0>a_m$ also have two critical points
 (Figure~\ref{fg8}\,$(b)$), but here the supercritical flows (curves 1 and 2)
 are bounded on one side, and the subcritical flow (curve 3) is bounded on
 both.
%
 \begin{figure}[!h]
 \epsfysize=100mm
 \centerline{\epsfbox{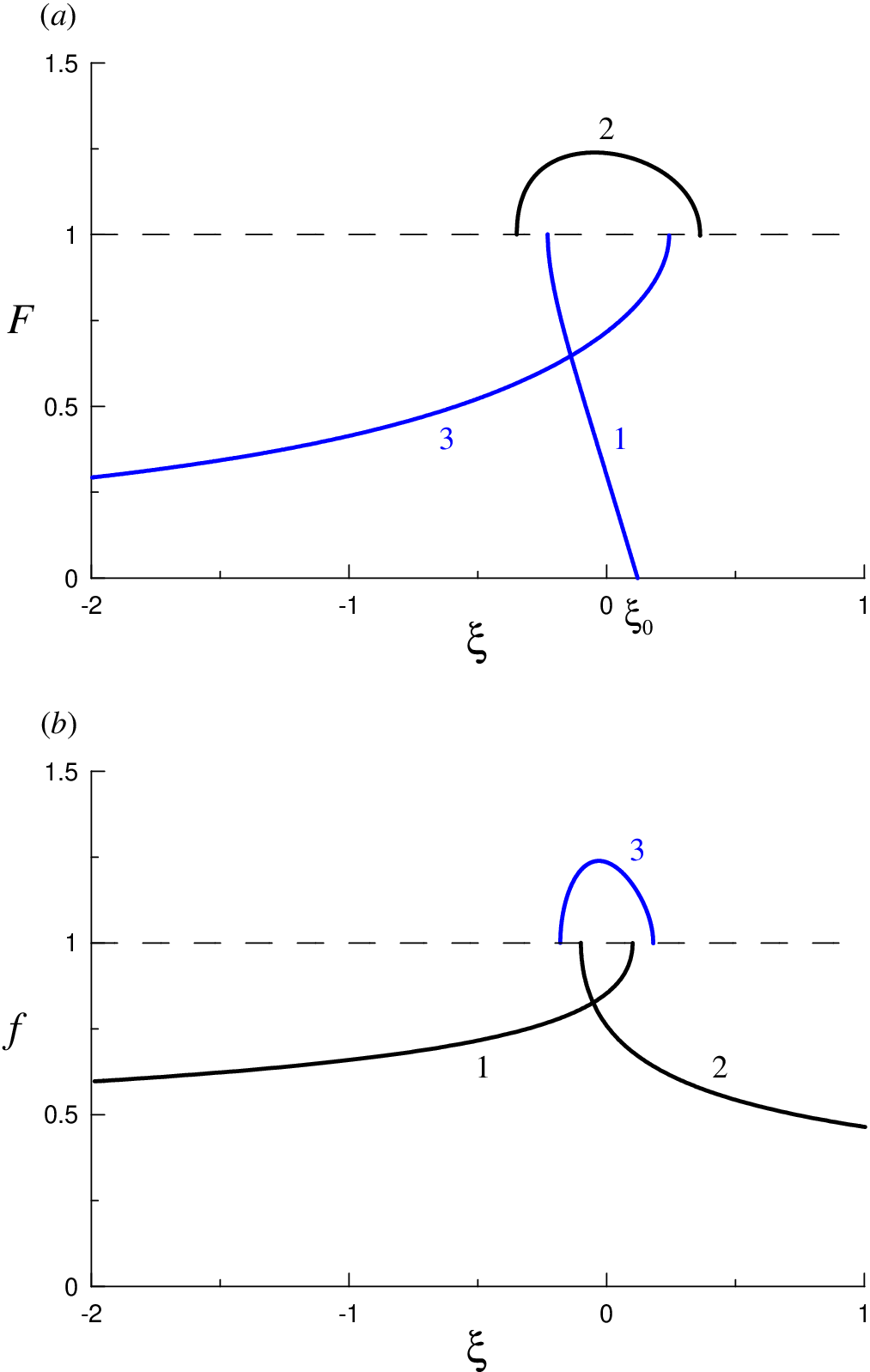}}
 \vspace{.2cm}
 \caption{Currents with two critical points: $(a)$ -- flows of class
 $\hB_+$, lines 1 and 3 correspond to subcritical branches, line 2 -- to
 supercritical one; $(b)$ -- flows of class $\hC_+$, lines 1 and 2 correspond
 to supercritical branches, line 2 -- to subcritical one. In both panels,
 $\tau_*=0.4$ and $c_0=1$.}
 \label{fg8}
 \end{figure}
 \subsubsection{Global flows}
 \label{sec4-12}
 \hspace\parindent
 The structure of the solution (\ref{Sol}) dictates a somewhat different
 approach to the existence of global flows than that adopted in
 Sections~\ref{sec3-22} and \ref{sec3-23}. The domain of definition of each
 flow described by such a solution lies in one of three intervals of $\tau$
 and is bounded by singular points. The flow is global if the transformation
 (\ref{xtau1}) maps these boundary points on the $\tau$-axis into infinitely
 distant points on the $\xi$-axis. An example is the flow of class $\hC_+$
 in Figure~\ref{fg7}\,$(b)$, for which Eq.~(\ref{xtau1}) maps $\tau=-\infty$
 into $\xi=+\infty$ and the boundary point $\tau=0$, in which $F\to\infty$,
 into $\xi=-\infty$ (see Eqs.~(\ref{as1C+}) and (\ref{as2C+})). In such cases,
 when integrating Eq.~(\ref{xtau1}), the integral over $\tau$ diverges at
 the boundary point, and it is sufficient that $c(\xi)$ does not have zero
 for finite $\xi$.

 If $a\to 0$ for finite $\tau$ and/or we need to avoid reaching a critical
 point at finite $\xi$, different requirements are needed to be imposed on
 $c(\xi)$. Because the integral over $\tau$ converges, the integral over
 $\xi$ must converge as well as $\xi\to+\infty$ ($\xi\to-\infty$). As can
 be seen from the first Eq.~(\ref{xtau1}), when $a\to 0$, $c(\xi)$
 should grow fast enough with $|\xi|$, so that $I_{F+}$ ($I_{F-}$)
 converges. In the case of a critical point, the condition is the same for
 subcritical ($F<1$) flow. If the flow is supercritical, it follows from the
 second Eq.~(\ref{xtau1}) that $c(\xi)$ should decrease when
 $|\xi|\to\infty$, so that the integral $I_{f+}$ ($I_{f-}$) converges. These
 conditions for the existence of global non-reflecting flows coincide with
 those obtained above for rank 1 flows (see Section~\ref{sec3-2}), as well
 as with those found previously for flows of classes $B$ (Eqs.~(I.5.5) and
 (I.5.6)) and $C$ (Eqs.~(II.3.15) and (II.3.26)). We shall return to
 this issue in Section~\ref{sec5}.
 \subsection{Singular solutions of equations (\ref{a2}) and (\ref{b2})}
 \label{sec4-2}
%
%
 \subsubsection{Equation (\ref{a2})}
 \label{sec4-21}
 \hspace\parindent
 Equation (\ref{a2}) has two obvious singular solutions that cannot be
 obtained from the general solution (\ref{Sol}) by choosing the constants
 $C$ and $\tau_*$,
 \be
 a(s) = \dfrac{1}{B_1(s_B-s)}\,,\ \ s_B=s_a-1,\ \ \quad \mbox{and}
 \quad a(s) = \dfrac{3}{B_1(s-s_a)^2}\,,\ \ B_1>0.
 \label{a2B}
 \ee
 The first of these satisfies the equation $a'=B_1a^2$ and describes the
 flows of class $B$ studied in Paper I. Therefore, the set $\hB$ of all
 solutions of the Eq.~(\ref{a2}) can be reasonably considered as an
 extension of class $B$. In particular, the flows of classes $\hB_+$ and
 $\hB_-$ considered above are subsets of class $\hB$.

 The second solution (\ref{a2B}) is an even function of $(s-s_a)$ singular at
 $s=s_a$. For $s<s_a$ and $s>s_a$, the flows it describes differ only in the
 change in the direction of the $x$ axis, so we restrict ourselves to the
 $s<s_a$ domain. It is easily seen that the transformation (\ref{sx}) maps
 both its ends into $x=+\infty$ ($c_\infty = \lim_{x\to+\infty}c(x)$):
 \be
 s\to-\infty:\ x \sim -c_\infty s,\ \ a \sim x^{-2},\quad \mbox{and} \quad
 s\to s_a-0:\ x \sim (s_a-s)^{-7},\ \ a \sim x^{2/7},
 \label{as-B2}
 \ee
 therefore, all flows are unbounded on the right ($x\to+\infty$).

 If $c(x)=c_0=\mbox{const}$, the critical point $U=c_0$ is reached at finite
 $x$ for both sub- and supercritical flows (see Figure~\ref{fg9}($a$)).
 Global flows can exist if $c(x)$ is such that the critical point shifts to
 $x=-\infty$. For this, the integral $I_{F-}$ must converge for subcritical
 flows and the integral $I_{f-}$ -- for supercritical ones.

 For a more detailed analysis, let us derive the equation just for $a(x)$.
 Expressing $s-s_a<0$ in terms of $a$ and using Eq.~(\ref{sx}) for returning
 to the variable $x$, we obtain
 \be
 \dfrac{\dd a}{\dd x} = 2\sqrt{\dfrac{B_1}{3}}\,\dfrac{a^{3/2}c}{U^2-c^2}\,,
 \quad \mbox{or} \quad \dfrac{\dd a^{-1/2}}{\dd\xi_a} = \dfrac{c}{c^2-U^2}\,,
 \quad \xi_a = \sqrt{\dfrac{B_1}{3}}\,x.
 \label{Ba}
 \ee

 If $c=c_0=\mbox{const}$, we arrive at the algebraic equation
 \be
 a^{7/2} + 7c_0^4a^{-1/2} = 7(\xi_a-\xi_{a0}), \quad \xi_{a0}=\mbox{const},
 \label{Ba0}
 \ee
 the left-hand side of which has a minimum at the critical point $a=c_0$
 ($\xi_a=\xi_{a0}+\frac{8}{7}\,c_0^{7/2}$). The plot $a(\xi_a)$ is
 shown in Figure~\ref{fg9}\,($a$).
%
 \begin{figure}[!h]
 \epsfxsize=80mm
 \centerline{\epsfbox{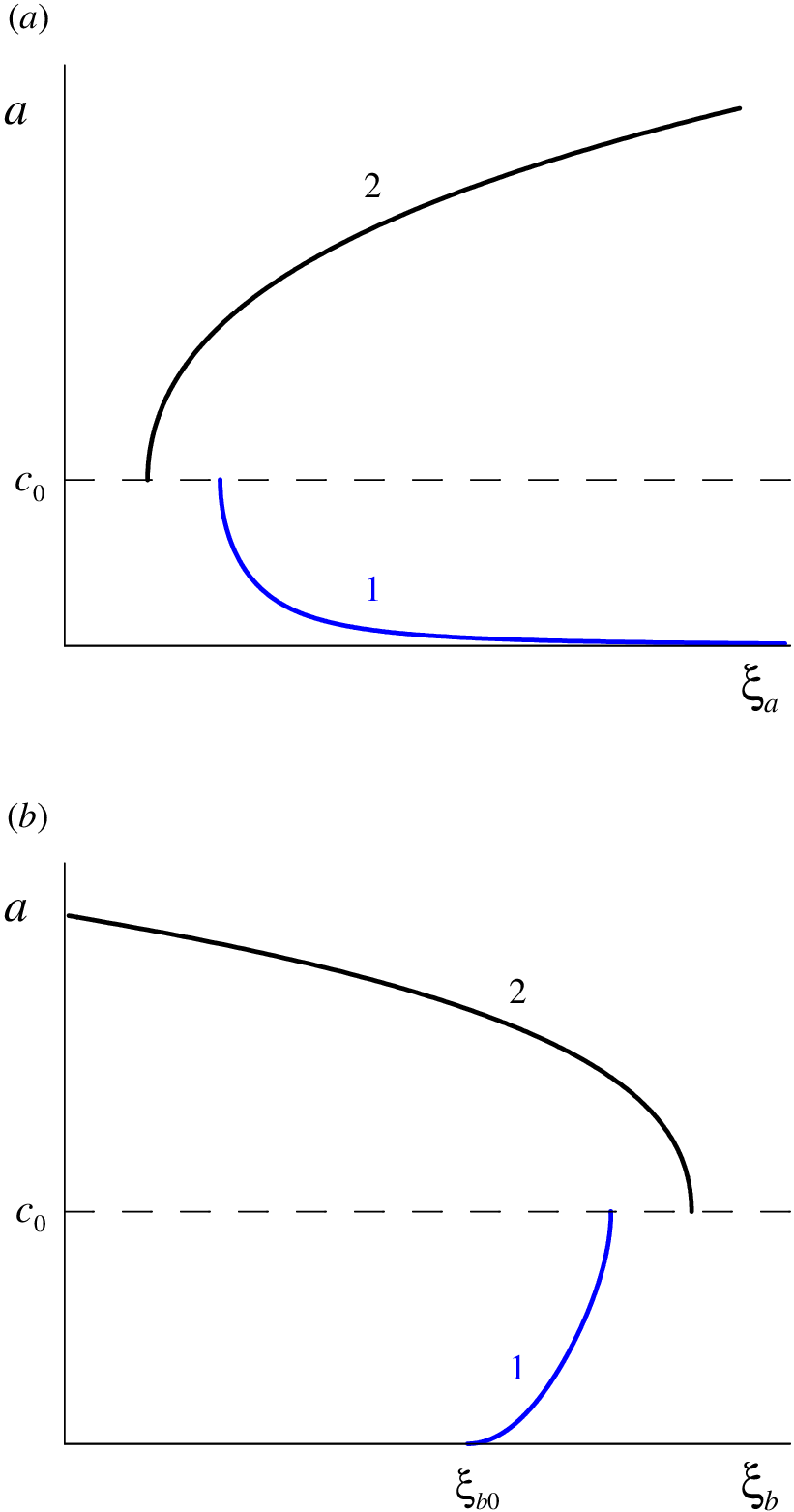}}
 \vspace{.2cm}
 \caption{Solutions ($a$) to Eq.~(\ref{Ba0}) and ($b$) to Eq.~(\ref{Ca0}).
 Curves 1 correspond to subcritical branches, and curves 2 -- to
 supercritical ones.}
 \label{fg9}
 \end{figure}

 To consider the global flows, it is convenient to introduce the functions
 \[
 G(\xi_a) = \left[\dfrac{a(\xi_a)}{c(\xi_a)}\right]^{1/4} \quad
 \mbox{and} \quad  g(\xi_a) = \dfrac{1}{G(\xi_a)}
 \]
 and represent the second Eq.~(\ref{Ba}) in several equivalent forms,
 \bea
 \label{Bsub}
 \dfrac{\dd a^{-1/2}}{\dd\xi_a} & = & \dfrac{1}{c(1-G^8)}\,,
  \\
 \label{Bsub1}
 c^{1/2}\dfrac{\dd G}{\dd\xi_a} & = & -\dfrac{G^2}{1-G^8} + M_aG, \quad
 M_a(\xi_a) = -\dfrac{\dd c^{1/2}}{\dd\xi_a}\,,
  \\
 \dfrac{\dd a^{7/2}}{\dd\xi_a} & = & \dfrac{7c^3}{1-g^8}\,.
 \label{Bsup}
 \eea

 {\bf Subcritical flows.} Integrating Eq.~(\ref{Bsub}),
 \[
 a^{-1/2}(\xi_a) = a^{-1/2}(\xi_{a1}) + \!\din_{\xi_{a1}}^{\xi_a}\!
 \dfrac{\dd y}{c(y)\Bl[1-G^8(y)\Br]}\,,
 \]
 we see that the trajectories can be extended to $\xi_a\to+\infty$ since
 $c(\xi_a)>0$. And for $\xi_a\to-\infty$, the convergence of
 $I_{F-}(\xi_{a1})$ is required. If the right-hand side has a positive limit
 as $\xi_a\to-\infty$, $a\to a_0>0$ and $U\sim D\sim c^{-1}$. However, in
 subcritical flows, an unlimited growth of $a$ and $c$ for $\xi_a\to-\infty$
 is possible as well. Then the limit of the left-hand side is zero, and if
 $c(\xi_a)\sim(-\xi_a)^{-\beta}$, $\beta>1$, we obtain assuming $F<1$
 everywhere,
 \[
 a^{-1/2} \sim (-\xi_a)^{1-\beta}\ \ \Longrightarrow\ \
 a \sim (-\xi_a)^{2(\beta-1)}, \quad F \sim (-\xi_a)^{\beta-2}.
 \]

 This estimate implies that there may be no global trajectories for $\beta>2$.
 To clarify this issue, let us turn to Eq.~(\ref{Bsub1}) and consider
 its NI $G=G_N(\xi_a)$,
 \[
 1 - G_N^8(\xi_a) = \dfrac{G_N(\xi_a)}{M_a(\xi_a)}\,.
 \]
 For every $M_a>0$ there is an unique solution that increases from zero for
 $M_a=0$ to 1 for $M_a\to+\infty$. If $\beta\ge 2$, $G_N(\xi)$ has a positive
 limit for $\xi_a\to-\infty$, and any trajectory passing below NI for a
 sufficiently large negative $\xi_a$ is global. Therefore, the convergence
 of $I_{F-}(\xi_{a1})$ is sufficient for the existence of global trajectories.

 {\bf Supercritical flows.} Integrating Eq.~(\ref{Bsup}),
 \[
 a^{7/2}(\xi_a) = a^{7/2}(\xi_{a1}) + 7\!\din_{\xi_{a1}}^{\xi_a}\!
 \dfrac{\dd y\,c^3(y)}{1-g^8(y)}\,,
 \]
 we see that for the existence of global trajectories, it is sufficient that
 $I_{f-}(\xi_{a1})$ converges. Indeed, one can take $a(\xi_{a1})$
 so large that the limit of the right-hand side is positive. Then
 \[
 a \to a_0>0, \quad g \to 0, \quad U \sim D \sim c^{-1} \to \infty.
 \]
 But even if $a$ and $c$ both tend to zero as $\xi_a\to-\infty$, then,
 for $c\sim(-\xi_a)^{-\beta}$, $\beta>1/3$,
 \[
 a \sim (-\xi_a)^{2(1-3\beta)/7}, \quad g \sim (-\xi_a)^{-(\beta+2)/14}, \quad
 U \sim (-\xi_a)^{(4-5\beta)/7}, \quad D \sim (-\xi_a)^{(19\beta-4)/7}.
 \]
 \subsubsection{Equation (\ref{b2})}
 \label{sec4-22}
 \hspace\parindent
 Equation (\ref{b2}) also has two singular solutions,
 \be
 a(s) \equiv \dfrac{1}{b(s)} = B_2(s_C-s),\ \ s_C=s_b-1,\ \ \quad
 \mbox{and} \quad a(s) = \dfrac{B_2}{3}\,\Bl(s-s_b\Br)^2,\ \ B_2>0,
 \label{a2C}
 \ee
 the first of which describes the flows of class $C$ described in Paper II.
 Therefore, the entire set of solutions to Eq.~(\ref{b2}), including
 the classes $\hC_\pm$, will be called the class $\hC$.

 The second solution (\ref{a2C}) is an even function of $(s-s_b)$, and
 $a(s_b)=0$. In the $s<s_b$ domain, this solution obeys the equation
 \be
 \dfrac{\dd a^{1/2}}{\dd\xi_b} = \dfrac{c}{c^2-U^2}\,,
 \quad \xi_b = \sqrt{\dfrac{B_2}{3}}\,x.
 \label{Ca}
 \ee

 The transformation (\ref{sx}) maps $s=-\infty$ into $x=-\infty$, but $s=s_b$
 and the critical point $U=c$ are generally mapped into finite values of $x$.
 Therefore, subcritical flows are usually defined over a finite interval of
 the $x$-axis. If the channel depth is constant ($c=c_0=\mbox{const}$), we
 arrive at the algebraic equation
 \be
 a^{9/2} - 9c_0^4a^{1/2} = 9(\xi_{b0}-\xi_b), \quad \xi_{b0}=\mbox{const}.
 \label{Ca0}
 \ee
 Its solution is shown in Figure~\ref{fg9}($b$) (cf. Figure~\ref{fg3}($c$)).
 When $\xi_b\to\xi_{b0}+0$, the subcritical branch tends to zero as
 $a\sim(\xi_b-\xi_{b0})^2/c_0^8$. The critical point is reached at
 $\xi_b=\xi_{b0}+\frac{8}{9}\,c_0^{9/2}$, and for $\xi_b\to-\infty$
 \[
 a \sim (-9\xi_b)^{2/9}, \quad U \sim D^{-1} \sim (-\xi_b)^{4/9}.
 \]

 To consider the existence of global solutions, we write Eq.~(\ref{Ca}) in
 several equivalent forms,
 \bea
 \label{Csub}
 \dfrac{\dd a^{1/2}}{\dd\xi_b} & = & \dfrac{1}{c(1-G^8)}\,,
  \\
 \label{Csub1}
 c^{3/2}\dfrac{\dd G}{\dd\xi_b} & = & \dfrac{1}{1-G^8} - M_bG, \quad
 M_b(\xi_b) = \dfrac{1}{3}\,\dfrac{\dd c^{3/2}}{\dd\xi_b}\,,
  \\
 \dfrac{\dd a^{9/2}}{\dd\xi_b} & = & -\dfrac{9c^3}{1-g^8}\,.
 \label{Csup}
 \eea

 {\bf Subcritical flows.} The flow is global if $c(\xi_b)$ is such that
 the transformation (\ref{sx}) maps $s=s_b$ into $x=-\infty$ and the critical
 point into $x=+\infty$. From Eq.~(\ref{Csub}) it follows that $a(\xi_b)$
 increases monotonically and
 \[
 a^{1/2}(\xi_b) = a^{1/2}(\xi_{b1}) + \!\din_{\xi_{b1}}^{\xi_b}\!
 \dfrac{\dd y}{c(y)\Bl[1-G^8(y)\Br]}\,, \quad \xi_{b1} = \mbox{const}.
 \]
 The trajectory is unbounded on the left if the integral $I_{F-}(\xi_{b1})$
 converges and the limit of the right-hand side as $\xi_a\to-\infty$ is
 positive or zero. In the latter case, for $c\sim(-\xi_b)^\beta$, $\beta>1$,
 \be
 a \sim |\xi_b|^{2(1-\beta)}, \quad G \sim |\xi_b|^{(2-3\beta)/2}, \quad
 U \sim |\xi_b|^{4-5\beta}, \quad D \sim |\xi_b|^{3\beta-4}.
 \label{as-C2}
 \ee

 For an unlimited continuation of the trajectory into the $\xi_b>\xi_{b1}$
 domain, the convergence of $I_{F+}(\xi_{b1})$ is sufficient, but not
 necessary. Indeed, let us consider NI of Eq.~(\ref{Csub1}),
 \[
 G_N^9(\xi_b) - G_N(\xi_b) = -M_b^{-1}(\xi_b).
 \]
 The left-hand side of this equation reaches its minimum value at
 $G_N=3^{-1/4}$, therefore, NI exists if $c(\xi_b)$ increases fast enough
 to satisfy the inequality
 \[
 M_b \ge M_{bc} = -\dfrac{1}{\min(G_N^9-G_N)} = \dfrac{3^{9/4}}{8}
 \approx 1.4806.
 \]
 As in Section~\ref{sec3-23}, NI is bounded on the left and has two branches,
 $G(\xi_b)=G_{N\,\pm}(\xi_b)$, between which $G(\xi_b)$ decreases. All
 trajectories intersecting the lower branch of the NI or lying below it
 are unbounded on the right. Moreover, $c(\xi_b)$ may grow more slowly
 than required for the convergence of $I_{F+}(\xi_{b1})$. For example,
 if $c(\xi_b)\sim\xi_b^\beta$ as $\xi_b\to+\infty$ and $2/3\le\beta<1$,
 formulas (\ref{as-C2}) hold. If $\beta<2/3$, the NI vanishes, and all
 trajectories reach $G=1$ for finite $\xi_b$. By and large the subcritical
 part of the phase portrait is very similar to that shown in Figure~\ref{fg5}.
 Global flows exist if $c(\xi_b)$ grows fast enough both for
 $\xi_b\to-\infty$ and for $\xi_b\to+\infty$.

 In {\bf supercritical flows} $a(\xi_b)$ decreases monotonically. Integrating
 Eq.~(\ref{Csup}),
 \[
 a^{9/2}(\xi_b) = a^{9/2}(\xi_{b1}) - 9\!\din_{\xi_{b1}}^{\xi_b}\!
 \dfrac{\dd y\,c^3(y)}{1-g^8(y)}\,,
 \]
 we see that for the existence of global trajectories, the convergence of
 $I_{f+}(\xi_{b1})$ is necessary and sufficient.
 \section{Physical discussion}
 \label{sec5}
 \hspace\parindent
 Let us start with main points of the factorization method. If the wave
 equation, after some change of dependent variables, can be factorized, that
 is, reduced to the form
 \[
 {\cal L}\left(\dfrac{\ptl g}{\ptl t} +
 w(x)\dfrac{\ptl g}{\ptl x}\right) = 0,
 \]
 where ${\cal L}$ is a linear operator, then one of its solutions is a wave
 of an arbitrary shape traveling with velocity $w(x)$,
 $g=\Psi\Bl(t-\int\dd x/w(x)\Br)$. In Papers I and II, it was shown that the
 substitution (see Eqs.~(\ref{Euler}) and (\ref{eta}))
 \[
 u(x,t) = \dfrac{\ptl\vp}{\ptl x}\ \ \ \mbox{and}\ \ \
 D(x)\zeta(x,t) = \dfrac{\ptl\phi}{\ptl x}
 \]
 provides wave equations for $\vp$ and $\phi$, each of which can be
 factorized under appropriate conditions on the flow parameters and has
 solutions of the form (\ref{TW1}). In the case of rank 0 flows (class $A$),
 that is, under the condition $a(x)=\mbox{const}$, both equations are
 factorized at once. In addition, each of them can also be factorized under
 its own conditions on $a(x)$, which define flows of classes $B$ and $C$
 constituting subsets of the rank 1 flows. Obviously, more complex
 substitutions can also lead to factorization. Finding them is difficult,
 but the theory \cite{Imsh,Goursat,Darboux} suggests an alternative way.
 If the wave equation can be factorized, then its solution in the form of a
 traveling wave (in terms of the physical variables $u$ and $\zeta$\,!)
 should be represented by a sum similar to Eq.~(\ref{TWF}). We use this way.

 Before going further, let us dwell on important property of the equations
 determining the behavior of non-reflecting flows. Equation (\ref{Eq-a1})
 for rank 1 flows is invariant under the simultaneous replacement $x\to -x$
 and $B_{1,2}\to -B_{1,2}$, whereas Eq.~(\ref{Eq-a2-1}) for rank 2 flows is
 invariant under replacement $x\to -x$ ($s\to -s$) at the same $B_{1,2}$.
 To some extent, this property erases the distinction between the ``upstream''
 and ``downstream'' directions. Indeed, if $U(x)>0$ and $c(x)>0$ satisfy
 the corresponding equation with $B_1=B_{10}$ and $B_2=B_{20}$, then $U(-x)$
 and $c(-x)$ satisfy the same equation with a proper choice of $B_{1,2}$.
 In other words, for a non-reflective flow with a certain upstream and
 downstream behavior, there is always a twin flow in which these properties
 are reversed. Therefore, in particular, when $\xi\to -\infty(+\infty)$ in
 Eqs.~(\ref{a1-}), (\ref{a1+}), (\ref{Ba}), and (\ref{Ca}), as well as in
 Figures \ref{fg3}, \ref{fg5}, and \ref{fg7} -- \ref{fg9}, it should not be
 understood as ``upstream'' (``downstream'').

 From an application perspective, the best properties are those of rank 0
 flows. Their depth and flow velocity are determined by the arbitrarily
 varying channel width $D$, see Eq.~(\ref{UHD}) and Figure~\ref{fg2}. In
 water at rest ($U=0$), their analog is represented by so-called
 self-consistent channels \cite{Pelin17a} whose width and depth are related
 in the same manner. These
 flows are defined along the entire $x$-axis, and their parameters can have
 virtually any desired properties (smoothness, monotonicity, boundedness,
 etc.) allowing them to remain within the model's applicability limits
 (shallow water, linearity, and lack of dissipation).

 As for flows of higher ranks, equations describing them have singularities.
 Therefore, these flows are typically defined either on a finite interval
 (curves 1 in Figures~\ref{fg3}\,(a,c) and \ref{fg9}\,(b)) bounded by the
 singularities of the solution, or on a half-line bounded by a singularity
 as well. And only in special cases they are global, that is, formally exist
 on the entire $x$-axis. In practice, we are usually interested in wave
 propagation over sufficiently large distances in preferred direction (e.g.,
 to some coastal structures). With this in mind, let us consider global flows
 and flows on a half-line.

 We are studying linear waves in shallow water flows, so the flow depth
 should remain finite, that is, neither too large nor too small. Then the
 flow behavior will be qualitatively the same as in the case of a constant
 depth, leaving almost no room for global flows. Of all the flows studied
 in the present paper and in Papers I and II, only supercritical flows of
 rank 1, similar to that shown by curve 2+ in Figure~\ref{fg3}\,(a) (see
 also the last paragraph of Section~\ref{sec3-22}), and of rank 2, similar
 to that shown in Figure~\ref{fg7}\,(b), can be considered as global flows.
 But even these flows have a shortcoming common for supercritical flows.
 Asymptotically, their velocities increase indefinitely (albeit slowly) in
 a narrowing channel, when $\xi\to-\infty$ in rank 1 flows (see Eq.~(\ref{as-a})),
 and in both directions in rank 2 flows, (see Eqs.~(\ref{as1C+}) and
 (\ref{as2C+})). Although formally such growth does not violate the
 approximations made, in practice it provokes the turbulence development.

 The existence of subcritical global solutions requires either the convergence
 of the integral $\int\dd\xi/c(\xi)$ as $\xi\to+\infty$ or $\xi\to-\infty$,
 or the occurrence of a null-isocline (see Sections~\ref{sec3-23} and II.3.2,
 as well as Figures~\ref{fg5}\,(a) and II.4). In any case, this implies a
 rapid (at least linear) increase of depth with distance, which leads to a
 violation of the shallow water approximation and makes necessary taking
 dispersion into account. Dispersion will affect the wave shape along with
 flow inhomogeneity, and it would be interesting to track their combined
 effects. Of particular interest is the question of whether dispersion can
 lead to a wave reflection. But this issue requires a separate in-depth study.

 In global supercritical solutions different from those discussed above, the
 integral $\int\dd\xi c^3(\xi)$ convergence as $\xi\to+\infty$ is required.
 This means that flow depth approaches zero, wave amplitude grows, and
 nonlinear processes come into play leading to a wave front steepening, wave
 breaking, bora formation, and other interesting and practically important
 phenomena. These issues are the subject of extensive literature, a review
 of which is beyond the scope of the paper. As both waves in supercritical
 flows propagate downstream, one might consider flows with depth decreasing
 upstream and start with linear waves. Unfortunately, the velocity in such
 flows should grow indefinitely downstream in a narrowing channel.
 Thus, one can see that global
 non-reflecting solutions are of theoretical rather than practical interest.

 The situation is somewhat better with solutions on a half-line. If the
 depth is everywhere finite, the flow behavior is qualitatively the same
 as for a constant depth. For subcritical flows, the domain of existence is
 limited by a singularity $U=c$ or $U=0$. These are the class $B$ flows
 studied in Paper I (see Figure~I.3\,(a)) and the class $\hB$ flows examined
 in Section~4 (Figures~\ref{fg7}\,(a), \ref{fg8}\,(a), curve 3, and
 \ref{fg9}\,(a), curve 1). Asymptotically, the velocity in them decreases as
 $U\sim|\xi|^{-2/3}$ with a proportional increase in the channel width $D$.
 Such flows can describe the transition between a channel or strait and
 the open sea. Solutions of the Eq.~(\ref{a1-}) tending to $a=1$ as
 $\xi\to-\infty$ (trajectories 1+ and $1-$ in Figure~\ref{fg3}\,(b)) are
 also practically feasible. In such flows, $U$ and $D$ remain finite
 along the entire half-line.

 As for the supercritical case, only solutions of Eq.~(\ref{a1-}) similar
 to 2+ in Figure~\ref{fg3}\,(a) could be classified as regular ``flows on
 a half-line'', although for $\xi\to-\infty$ their practical applicability
 is limited by the increase in velocity rather than singular point.

 If the depth $H$ is everywhere finite, the critical point $U=c$ limits the
 existence domain of any of the flows examined, except for similar to those
 shown in Figure~\ref{fg7}, which do not reach it, and flows of rank 0 (class
 $A$), for which this point is not singular. Unruh's seminal paper
 \cite{Unruh81} initiated intensive study of laminar transcritical flows
 \cite{ArtBH,AnGrav}, with primary emphasis upon radiation, transmission,
 reflection, and amplification of waves. Today, such flows have been
 thoroughly studied both theoretically \cite{SchUn,AnalogGrav} and in
 laboratory experiments \cite{Wein1,Wein2}.

 Let us dwell on the behavior of waves in the process of a rank 0 flow
 transition through a critical point. In the downstream direction, such a
 process can be twofold, with the transition from subcritical to
 supercritical flow (Sub-Super), or vice versa (Super-Sub). The accelerated
 wave propagates freely downstream, ``ignoring'' any critical points, and
 its amplitude changes in accordance with Eq.~(\ref{form0}), top row. As
 for the retarded wave, let the function $F_2$ (see Eq.~(\ref{-})) to be
 initially localized in a narrow domain far from the critical points. The
 wave travels upstream in subcritical parts of the flow and downstream
 in its supercritical parts. Hence, in any case, it can not reach a
 Sub-Super transition point but travels invariably toward a Super-Sub one.
 As it approaches this point, its wavelength tends to zero, and the
 wave is eventually absorbed even at low viscosity. For clarity sake, let
 us make some estimates. To do this, we supplement Eq.~(\ref{Euler})
 with nonlinear and viscous terms and evaluate their contributions,
 \[
 \underbrace{\dfrac{\ptl u}{\ptl t}}_{\vep U_c^2/l}\ +\
 \underbrace{\dfrac{\ptl(Uu)}{\ptl x} + \dfrac{\ptl\zeta}{\ptl x}}_
 {\vep U_c^2 L/l^2}\ +
 \underbrace{u\,\dfrac{\ptl u}{\ptl x}}_{\vep^2U_c^2 L^2/l^3} =
 \underbrace{\nu\,\dfrac{\ptl^2 u}{\ptl x^2}}_{\nu\vep U_c L/l^3}\,,
 \]
 taking into account the behavior of form-factors near the critical point
 (see the bottom row of Eq.~(\ref{form0})). Here, $U_c$ is the flow
 velocity at the critical point, $\vep=O(ul/LU_c)$ is the relative disturbance
 amplitude, $L$ and $l$ are the initial and current disturbance scales, and
 $\nu$ is the kinematic viscosity. As the characteristic wavelength $l$
 decreases, viscosity comes into play and absorbs the wave when $l=O(\nu/U_c)$.
 On the other hand, the problem remains linear as far as $\vep\ll l/L$, or
 $\vep\ll R^{-1}$, where $R=LU_c/\nu$ is the Reynolds number. In more detail,
 this problem is analyzed in \cite{ChErSt,ChSt21}.

 Finally, let us consider one more issue. Wave propagation without reflection
 in a inhomogeneous flow undoubtedly requires a precise adjustment of flow
 parameters. The goal of this paper and other similar studies is to find
 suitable combinations of them. Clearly, if the flow parameters deviate from
 ``ideality'', waves will be reflected. Based on the general concept of a
 ``continuous dependence on the parameters'', it should be expected that,
 far from singular points, a small deviation will result in weak reflection.
 Another interesting problem is the wave propagation in flows composed of
 non-reflecting fragments. It was shown (e.g., \cite{Diden08}) that,
 as a rule, the waves are markedly reflected on fragment boundaries. The
 exception is provided by some flows in which adjacent fragments satisfy the
 same compatibility condition (that is, the same equation for $a(x)$, see
 \cite{ChSt22}, Section~6.1). And the intriguing question of what happens
 to waves when passing through a fold-type singularity at $U=c$ remains
 unexplored.
 \section{Conclusions}
 \label{sec6}
 \hspace\parindent
 If linear waves in a stationary inhomogeneous flow are described by a
 second-order hyperbolic equation (or system of equations), the Laplace
 cascade method serves as the basis for the algorithm described in
 Section~\ref{sec2}, which is exhaustive for finding non-reflecting flow
 configurations. Without going into the details of the wave equation
 transformations, its solutions for each of the traveling waves should be
 sought as a sum (\ref{TWF}). For surface waves on a shallow water flow with
 $x$-dependent parameters, we have derived a system of equations expressing
 the form-factors contained in the sum in terms of the flow parameters for
 an arbitrary number of summands $n$. Solving these equations requires a
 compatibility condition to be fulfilled, which is an ordinary differential
 equation that $a(x)$ should obey for the waves can propagate without
 reflection. The equation depends on $n$, so the flows described by it are
 called the rank $n$ flows.

 Flows of rank 0 and a part of rank 1 flows have been studied in Papers
 I and II. We have derived a general equation (\ref{Eq-a1}) describing the
 structure of all rank 1 flows and studied those of its solutions that could
 not be found by previously used methods. Among them, the most interesting
 are the flows for which the rank 0 flow with the same constants $B_{1,2}$
 serves as an attractor (see Figure~\ref{fg3}($a,\,b$)). They are regular
 on a semi-line and therefore are of much interest for applications.

 In addition, we have derived a compatibility condition for rank 2 flows.
 It is the nonlinear third-order differential equation (\ref{R2}) (in a more
 easily understood form, Eq.~(\ref{Eq-a2-3})). It turned out to be quite
 complex, and it was possible to study only those of its solutions that
 correspond to the zero value of one of the constants $B_{1,2}$. As was
 shown, their properties are in the main similar to those characteristic of
 the rank 1 flows under the same condition. This allows us to speak of
 extensions $\hB$ and $\hC$ of the $B$ and $C$ classes of flows. However,
 some of newly found solutions have properties different from those
 inherent to the flows of $B$ and $C$ classes studied in Papers I and II.
 For example, some flows of finite depth do never reach the critical point
 (see Figire~\ref{fg7}).

 Considerable attention was devoted to finding global flows
 defined along the entire $x$-axis. Unfortunately, with the exception of
 some cases (such as supercritical flows shown in Figures~\ref{fg3}($a$) and
 \ref{fg7}($b$)), their existence is determined by special requirements on
 the flow parameters, which, even at finite $x$, make necessary going beyond
 the applicability limits of the flow model (for more details, see
 Section~\ref{sec5}).
 \medskip\newline
 {\bf Acknowledgements.} The author is thankful to Professors O.V. Kaptsov
 and S.P. Tsarev for helpful discussions on the Laplace cascade method.
 \medskip\newline
 {\bf Research ethics}: Not applicable.
 \medskip\newline
 {\bf Informed consent}: Not applicable.
 \medskip\newline
 {\bf Author contributions}: The author has accepted responsibility for the
 entire content of this manuscript and approved its submission.
 \medskip\newline
 {\bf Use of Large Language Models, AI and Machine Learning Tools}: None
 declared.
 \medskip\newline
 {\bf Conflict of interest}: The author states no conflict of interest.
 \medskip\newline
 {\bf Research funding}: The work was financially supported by the Ministry
 of Science and Higher Education of the Russian Federation.
 \medskip\newline
 {\bf Data availability}: Not applicable.
 \medskip\newline
 {\bf Author ORCID.} https://orcid.org/0000-0001-5543-2474.
 \appendix
 \section{The Laplace cascade method}
 \label{AppA}
 \hspace\parindent
 To reduce equations (\ref{Euler}) and (\ref{eta}) to a characteristic form,
 let us introduce differentiation operators along the characteristics
 (\ref{har}),
 $$
 \cD_{1,2} = \dfrac{\ptl}{\ptl t} + w_{1,2}(x)\,\dfrac{\ptl}{\ptl x},
 \qquad \cD_1\cD_2 - \cD_2\cD_1 = G_0(x)\Bl(\cD_1 - \cD_2\Br),
 \eqno{(\rm A1)}
 $$
 where $w_{1,2}(x)=U(x)\pm c(x)$,
 \[
 G_0(x) = \dfrac{w_1(x)w'_2(x)-w'_1(x)w_2(x)}{w_1(x)-w_2(x)} \equiv
 U(x)\left[\dfrac{U'(x)}{U(x)} - \dfrac{c'(x)}{c(x)}\right],
 \]
 and the prime denotes the derivative with respect to $x$. The functions
 $v_\pm(x,t)=\zeta(x,t)\pm c(x)\,u(x,t)$ obey the equations
 $$
 \Bl[\cD_1 + \al_1(x)\Br]v_+(x,t) = \beta_1(x)v_-(x,t),
 \eqno{(\rm A2)}
 $$
 $$
 \Bl[\cD_2 + \al_2(x)\Br]v_-(x,t) = \beta_2(x)v_+(x,t),
 \eqno{(\rm A3)}
 $$
 where
 \[
 \ba{l}
 \al_1 = \dfrac{1}{2}\left[(U-c)\left(\dfrac{U'}{U} - \dfrac{c'}{c}\right)
 - 2(U+c)\dfrac{c'}{c}\right], \quad \beta_1 = (U-c)\dfrac{\dd\ln a}{\dd x}\,,
   \\ \\
 \al_2 = \dfrac{1}{2}\left[(U+c)\left(\dfrac{U'}{U} - \dfrac{c'}{c}\right)
 - 2(U-c)\dfrac{c'}{c}\right], \quad \beta_2 = (U+c)\dfrac{\dd\ln a}{\dd x}\,.
 \ea
 \]
 If (see (\ref{R0})) $a^2(x)\equiv c(x)U(x)=\mbox{const}$, then
 $\beta_1=\beta_2=0$, and $v_+$ and $v_-$ describe the accelerated and
 retarded waves, respectively,
 \[
 v_+(x,t) = \dfrac{\Phi_1(T_+)}{U^2(x)+a^2}\,, \quad
 v_-(x,t) = \dfrac{\Phi_2(T_-)}{U^2(x)-a^2}\,,
 \]
 where $\Phi_{1,2}(X)$ are arbitrary functions.

 Equations (A2) and (A3) can be reduced to a single
 equation for $v_+$ if $\beta_1\ne 0$,
 \[
 \Bl[\cD_2 + \al_2 - w_2(\ln|\beta_1|)'\Br]\Bl(\cD_1 + \al_1\Br)v_+ =
 \beta_1\beta_2v_+,
 \]
 and to a single equation for $v_-$ when $\beta_2\ne 0$,
 \[
 \Bl[\cD_1 + \al_1 - w_1(\ln|\beta_2|)'\Br]\Bl(\cD_2 + \al_2\Br)v_- =
 \beta_1\beta_2v_-.
 \]
 Clearly, these equations are equivalent, and we focus on the last one.
 Using the relations (A1), it can be written in two forms,
 $$
 \Bl[\cD_1 + A_0(x)\Br]\Bl[\cD_2 + B_0(x)\Br]F_0(x,t) =
 h_0(x)F_0(x,t)
 \eqno{(\rm A4)}
 $$
 and
 $$
 \Bl[\cD_2 + \bB_0(x)\Br]\Bl[\cD_1 + \bA_0(x)\Br]F_0(x,t) =
 k_0(x)F_0(x,t),
 \eqno{(\rm A5)}
 $$
 where $F_0(x,t) \equiv v_-(x,t)$ and
 \[
 \ba{l}
 A_0(x) = \al_1(x) - w_1(x)(\ln|\beta_2(x)|)',
 \quad B_0(x) = \al_2(x), \quad h_0(x) = \beta_1(x)\beta_2(x),
   \\ \\
 \bA_0(x) = A_0(x) - G_0(x), \quad \bB_0(x) = B_0(x) + G_0(x),
   \\ \\
 k_0(x) = h_0(x) + w_2(x)\bA'_0(x) - w_1(x)B'_0(x) +
 [\bA_0(x)-\bB_0(x)]G_0(x) + G_0^2(x).
 \ea
 \]
 The functions $h_0(x)$ and $k_0(x)$ are called the Laplace invariants of
 equations (A4) and (A5). We assume that they are nonzero and
 proceed to cascade transformations.

 {\bf $\cD_1$--transformations.} Let $F_1(x,t)=\Bl[\cD_2 +
 B_0(x)\Br]F_0(x,t)$. Then, according to (A4),
 $F_0(x,t)=h_0^{-1}(x)\Bl[\cD_1+A_0(x)\Br]F_1(x,t)$, and $F_1(x,t)$
 satisfies the equation
 \[
 \ba{l}
 \Bl[\cD_2 + \bB_1(x)\Br]\Bl[\cD_1 + \bA_1(x)\Br]F_1(x,t) = k_1(x)F_1(x,t),
   \\ \\
 \bA_1(x) = A_0(x), \quad \bB_1(x) = B_0(x) - w_2(x)\Bl(\ln|h_0(x)|\Br)',
 \quad k_1(x) = h_0(x).
 \ea
 \]
 Let us transform it to a form similar to (A4):
 \[
 \ba{l}
 \Bl[\cD_1 + A_1(x)\Br]\Bl[\cD_2 + B_1(x)\Br]F_1(x,t) = h_1(x)F_1(x,t),
   \\ \\
 A_1(x) = A_0(x) + G_0(x), \quad B_1(x) = B_0(x) - G_0(x) -
 w_2(x)\Bl(\ln|h_0(x)|\Br)',
   \\ \\
 h_1(x) = 2h_0(x) - k_0(x) - w_1(x)\Bl[w_2(x)\Bl(\ln|h_0(x)|\Br)'\Br]' -
 G_0(x)w_2(x)\Bl(\ln|h_0(x)|\Br)'
   \\ \\ \phantom{www}
 - \Bl[w_1(x)+w_2(x)\Br]G'_0(x) - 2G_0^2(x).
 \ea
 \]

 If $h_1(x)\ne 0$, we can introduce $F_2(x,t)$, and so on. At the $m$-th
 step, we obtain the equation
 $$
 \Bl[\cD_1 + A_m(x)\Br]\Bl[\cD_2 + B_m(x)\Br]F_m(x,t) = h_m(x)F_m(x,t),
 \eqno{(\rm A6)}
 $$
 where
 \[
 \ba{l}
 F_m(x,t) = \Bl[\cD_2 + B_{m-1}(x)\Br]F_{m-1}(x,t), \quad
 F_{m-1}(x,t) = h_{m-1}^{-1}(x)\Bl[\cD_1 + A_{m-1}(x)\Br]F_m(x,t),
   \\ \\
 A_m(x) = A_{m-1}(x) + G_0(x), \quad B_m(x) = B_{m-1}(x) - G_0(x) -
 w_2(x)\Bl(\ln|h_{m-1}(x)|\Br)',
   \\ \\
 h_m(x) = 2h_{m-1}(x) - k_{m-1}(x) - w_1(x)\Bl[w_2(x)\Bl(\ln|h_{m-1}(x)|
 \Br)'\Br]' - G_0(x)w_2(x)\Bl(\ln|h_{m-1}(x)|\Br)'
   \\ \\ \phantom{wwwi}
 - \Bl[w_1(x)+w_2(x)\Br]G'_0(x) - 2G_0^2(x), \quad k_m(x) = h_{m-1}(x).
 \ea
 \]
 If $h_m(x)=0$, equation (A6) factorizes and has a particular solution
 \[
 F_m^{(-)}(x,t) = \Phi_-(T_-)\exp\left[-\!\int\!
 \dfrac{\dd x\,B_m(x)}{U(x)-c(x)}\right]
 \]
 with arbitrary $\Phi_-(T_-)$. This is a retarded wave. Expressing
 $F_{m-1}^{(-)}(x,t)$ in terms of $F_m^{(-)}(x,t)$ and so on, we see that
 differentiation takes place at each step, so the corresponding solution
 to equation (A4) has the form (cf. equations (\ref{-}))
 $$
 F_0^{(-)}(x,t) = a_0^{(-)}(x)\Phi_-(T_-) +
 \sum\limits_{k=1}^{m}a_k^{(-)}(x)\Phi_-^{(k)}(T_-).
 \eqno{(\rm A7)}
 $$

 {\bf $\cD_2$--transformations.} Let $F_{-1}(x,t)=\Bl[\cD_1 +
 \bA_0(x)\Br]F_0(x,t)$. Then, according to (A5),
 $F_0(x,t)=k_0^{-1}(x)[\cD_2+\bB_0(x)]F_{-1}(x,t)$, and
 $F_{-1}(x,t)$ satisfies the equation
 \[
 \ba{l}
 \Bl[\cD_1 + A_{-1}(x)\Br]\Bl[\cD_2 + B_{-1}(x)\Br]F_{-1}(x,t) =
 h_{-1}(x)F_{-1}(x,t),
    \\ \\
 A_{-1}(x) = \bA_0(x) - w_1(x)\Bl(\ln|k_0(x)|\Br)', \quad
 B_{-1}(x) = \bB_0(x), \quad h_{-1}(x) = k_0(x).
 \ea
 \]
 Let us transform it into an equivalent equation
 \[
 \Bl[\cD_2 + \bB_{-1}(x)\Br]\Bl[\cD_1 + \bA_{-1}(x)\Br]F_{-1}(x,t)
 = k_{-1}(x)F_{-1}(x,t),
 \]
 where
 \[
 \ba{l}
 \bA_{-1}(x) = \bA_0(x) - G_0(x) - w_1(x)\Bl(\ln|k_0(x)|\Br)', \quad
 \bB_{-1}(x) = \bB_0(x) + G_0(x),
   \\ \\
 k_{-1}(x) = 2k_0(x) -  h_0(x) - w_2(x)\Bl[w_1(x)\Bl(\ln|k_0(x)|\Br)'\Br]' -
 G_0(x)w_1(x)\Bl(\ln|k_0(x)|\Br)'
   \\ \\ \phantom{wwwn}
 - \Bl[w_1(x) + w_2(x)\Br]G'_0(x) - 2G_0^2(x).
 \ea
 \]
 If $k_{-1}(x)\ne 0$, we introduce $F_{-2}(x,t)$, and so on. At the
 $n$-th step
 \[
 F_{-n+1}(x,t) = \dfrac{1}{k_{-n+1}(x)}\Bl[\cD_2+\bB_{-n+1}(x)\Br]
 F_{-n}(x,t), \quad h_{-n}(x) = k_{-n+1}(x),
 \]
 and
 $$
 \Bl[\cD_2 + \bB_{-n}(x)\Br]\Bl[\cD_1 + \bA_{-n}(x)\Br]F_{-n}(x,t)
 = k_{-n}(x)F_{-n}(x,t),
 \eqno{(\rm A8)}
 $$
 where
 \[
 \ba{l}
 \bA_{-n}(x) = \bA_{-n+1}(x) - G_0(x) - w_1(x)\Bl(\ln|k_{-n+1}(x)|\Br)', \quad
 \bB_{-n}(x) = \bB_{-n+1}(x) + G_0(x),
    \\ \\
 k_{-n}(x) = 2k_{-n+1}(x) - h_{-n+1}(x) - w_2(x)\Bl[w_1(x)\Bl(\ln|k_{-n+1}(x)|
 \Br)'\Br]'
    \\ \\ \phantom{wwwn}
 - G_0(x)w_1(x)\Bl(\ln|k_{-n+1}(x)|\Br)'
 - \Bl[w_1(x) + w_2(x)\Br]G'_0(x) - 2G_0^2(x).
 \ea
 \]

 If $k_{-n}(x)=0$, equation (A8) factorizes and has an obvious
 particular solution
 \[
 F_{-n}^{(+)}(x,t) = \Phi_+(T_+)\exp\left[-\!\int\!
 \dfrac{\dd x\bA_{-n}(x)}{U(x)+c(x)}\right]
 \]
 with arbitrary $\Phi_+(T_+)$. This is an accelerated wave, and the corresponding
 solution of equation (A4) has the form (cf. equations (\ref{+}))
 $$
 F_0^{(+)}(x,t) = a_0^{(+)}(x)\Phi_+(T_+) +
 \sum\limits_{k=1}^{n}a^{(+)}_k(x)\Phi_+^{(k)}(T_+).
 \eqno{(\rm A9)}
 $$
 We emphasize that equation (A4) has solutions of the form (A7)
 and (A9) if and only if $h_q(x)$ and $k_{-p}(x)$ vanish for some
 $q$ and $p$ such that $0\le q\le m$ and $0\le p\le n$, see
 \cite{Goursat,Darboux}.
 \section{Solution of equation (\ref{Eq-y})}
 \label{AppB}
 \hspace\parindent
 According to \cite{PZ03} (see equation 2.6.3.17), the solution to
 Eq.~(\ref{Eq-y}) can be expressed (after correcting a misprint) in terms
 of the parameter $\tau_P>0$ as
 $$
 s - s_0 =\dfrac{a}{C_1E_1\tau_P}\Bl(\tau_PF^2 + \tau_P^2F - E_1^2\Br),
 \quad y = \dfrac{bC_1^2}{F}\,, \quad B = -\dfrac{1}{a^2b},
 \eqno{(\rm B1)}
 $$
 where
 \[
 E_1(\tau_P) = \sqrt{\tau_P(1+\tau_P)} - \ln\Bl(\sqrt{\tau_P}
 +\sqrt{1+\tau_P}\Br) + C_2, \quad
 F(\tau_P) = E_1(\tau_P)\sqrt{\dfrac{1+\tau_P}{\tau_P}} - \tau_P,
 \]
 and $a$, $b$, $C_1$, and $C_2$ are constant. Direct calculations show that
 \[
 \ba{l}
 s -s_0 \equiv \dfrac{a(\tau_PF^2 + \tau_P^2F - E_1^2)}{C_1E_1\tau_P} =
 \dfrac{a}{C_1}\Bl[C_2 - \ln\Bl(\sqrt{\tau_P} + \sqrt{1+\tau_P}\Br)\Br],
    \\ \\
 F = 1 + \sqrt{\dfrac{1+\tau_P}{\tau_P}}\Bl[C_2 - \ln\Bl(\sqrt{\tau_P}
 +\sqrt{1+\tau_P}\Br)\Br].
 \ea
 \]
 It is convenient to introduce a new parameter $\tau$ by $\tau_P=\sinh^2\tau$.
 Then
 $$
 s - s_0 = \dfrac{a}{C_1}(C_2 - \tau), \quad
 F = 1 + (C_2 - \tau)\coth\tau, \quad y = \dfrac{bC_1^2}{F}\,.
 \eqno{(\rm B2)}
 $$

 Equation (\ref{Eq-y}) is invariant under the transformation
 \[
 s - s_0\ \to\ \lb(s - s_0), \quad y\ \to\ y/\lb^2, \quad
 \lb = \mbox{const},
 \]
 which corresponds to the change $a\ \to\ \lb a$ and $b\ \to\ b/\lb^2$ in
 Eqs.~(B1) and (B2). Without loss of generality, one can put $a=1$, then
 $bB=-1$. Denoting $C=C_1$, $\tau_*=C_2$, and $E(\tau)=-F(\tau)$, we
 arrive at Eq.~(\ref{Sol}).
}

\end{document}